\documentclass{aa}

\usepackage{hyperref}
\usepackage{bm}
\usepackage{amsmath}
\usepackage{subfigure}
\usepackage{color}
\usepackage{graphicx}
\usepackage[varg]{txfonts}
\usepackage{natbib}

\bibpunct{(}{)}{;}{a}{}{,}

\definecolor{dark-blue}{rgb}{0.0,0.0,0.745}

\begin{document}

\title{Toward Understanding the Origin of Asteroid Geometries: Variety in Shapes Produced by Equal-Mass Impacts}

%% include affiliations in footnotes:
\author{K. Sugiura
  \and H. Kobayashi
  \and S. Inutsuka}

\institute{Department of Physics, Nagoya University, Aichi 464-8602, Japan \email{sugiura.keisuke@a.mbox.nagoya-u.ac.jp}}

\date{Received / Accepted }

\abstract{More than a half of asteroids in the main belt have irregular shapes with the ratios of the minor to major axis lengths less than 0.6. One of the mechanisms to create such shapes is collisions between asteroids. The relationship between shapes of collisional outcomes and impact conditions such as impact velocities may provide information on the collisional environments and its evolutionary stages when those asteroids are created. In this study, we perform numerical simulations of collisional destruction of asteroids with radii $50\,{\rm km}$  and subsequent gravitational reaccumulation using Smoothed Particle Hydrodynamics for elastic dynamics with self-gravity, a model of fracture of rock, and a model of friction of completely damaged rock. We systematically vary the impact velocity from $50\,{\rm m/s}$ to $400\,{\rm m/s}$ and the impact angle from $5^{\circ}$ to $45^{\circ}$. We investigate shapes of the largest remnants resulting from collisional simulations. As a result, various shapes (bilobed, spherical, flat, elongated, and hemispherical shapes) are formed through equal-mass and low-velocity ($50 - 400\,{\rm m/s}$) impacts. We clarify a range of the impact angle and velocity to form each shape. Our results indicate that irregular shapes, especially flat shapes, of asteroids with diameters larger than $80\,{\rm km}$ are likely to be formed through similar-mass and low-velocity impacts, which are likely to occur in the primordial environment prior to the formation of Jupiter. }

\maketitle

\section{Introduction}
Planets are formed in protoplanetary disks around protostars through collisional coalescence of planetesimals (\citealt{Safronov1969,Hayashi-et-al1985}). The growth mode of planetesimals is considered as ``runaway'', that is, larger planetesimals grow more rapidly than smaller ones (e.g., \citealt{Greenberg-et-al1978,Wetherill-and-Stewart1989,Kokubo-and-Ida1996}). The runaway growth produces a bimodal mass distribution of bodies composed of protoplanets and remnant planetesimals of mass around the onset of runaway growth (\citealt{Kobayashi-et-al2016}). Main-belt asteroids located between the orbits of Mars and Jupiter may be remnants of planetesimals (e.g., \citealt{Petit-et-al2001,Bottke-et-al2005}). A large number of asteroids (more than one hundred thousand asteroids for those with diameters > $1\,{\rm km}$) allows statistical discussion to reveal the history of the solar system. 

Asteroids have variety of shapes. Recent in-situ observations by spacecraft and light curve and radar observations by ground-based telescopes reveal shapes of about 1,000 asteroids. Those obtained from light curves are summarized in Database of Asteroid Models from Inversion Techniques (DAMIT; \citealt{Durech-et-al2010}). According to asteroidal shapes obtained from various observations, shapes of many asteroids smaller than $100\,{\rm km}$ are distinctly different from planet shapes, which are almost spheres (\citealt{Fujiwara-et-al2006,Durech-et-al2010,Marchis-et-al2014,Cibulkova-et-al2016}). Some asteroids with diameters larger than $100\,{\rm km}$ also have irregular shapes. For example, (624) Hektor has a very elongated shape with the intermediate axis length of $195\,{\rm km}$ and the major axis length of $370\,{\rm km}$ (\citealt{Storrs-et-al1999}). In this paper, we define irregular shapes as shapes with axis ratios less than 0.6.

Irregular shapes of asteroids may be formed through collisional destruction of planetesimals. Irregular-shape formation of rubble piles through collisional destruction of planetesimals and gravitational reaccumulation is investigated using Smoothed Particle Hydrodynamics (SPH) method or N-body code with models of material strength. Some impact simulations reproduce formation of elongated shapes like (25143) Itokawa or bilobed shapes like (67P) Churyumov-Gerasimenko (\citealt{Michel-and-Richardson2013,Jutzi-and-Asphaug2015,Jutzi-and-Benz2017,Schwartz-et-al2018}). 

Shapes of objects formed through collisional destruction or coalescence depend on impact conditions (e.g., \citealt{Jutzi-and-Asphaug2015}). For example, collisions with impact velocities comparable to the escape velocity result in simply merging. In contrast, collisions with higher impact velocities result in catastrophic destruction, and shapes are determined through gravitational reaccumulation of fragments. Thus the relationship between impact conditions and resultant shapes clarifies impact conditions (e.g., impact velocities) to form irregular shapes of asteroids and then gives constraints on the collisional environment (e.g., eccentricity distribution) that forms asteroids.

Collisional lifetimes of asteroids with diameters $\ga 10\,{\rm km}$ in the present main belt are estimated to be $\sim 10\,{\rm Gyr}$, which is longer than the age of the solar system (\citealt{Obrien-and-Greenberg2005}). Although collisional destruction of asteroids with diameters $\ga 10\,{\rm km}$ in the present solar system is rare, destructive collisions are expected to be more frequent in the primordial environment because the formation and migration of Jupiter may significantly deplete asteroids (\citealt{Bottke-et-al2005,Walsh-et-al2011}). Thus the shapes of large asteroids may preserve those formed in the primordial solar system. In the planet formation era, eccentricities of planetesimals are small around young planets with small masses (e.g., \citealt{Wetherill-and-Stewart1993, Kokubo-and-Ida1998, Inaba-et-al2003, Kobayashi-and-Tanaka2018}), and impact velocities between young asteroids may be $ < 1\,{\rm km/s}$ because of the absence of large perturbers. Therefore investigations of shapes formed through impacts with low impact velocities may suggest when asteroidal shapes were formed (e.g., prior to or after Jupiter formation).

In this study, we perform impact simulations to investigate collisional destruction of planetesimals and subsequent gravitational reaccumulation using SPH method. For impacts with low impact velocities, the most efficient collisions to change shapes are expected to be equal-mass impacts because impacts with high mass ratio tend to result in partial deformation such as the formation of craters rather than catastrophic destruction. DAMIT already includes irregular shapes of $\sim 200$ asteroids with radii $\sim 50\,{\rm km}$. Therefore, we consider collisions between two planetesimals with the radius of $50\,{\rm km}$, and we focus on the dependence of resultant shapes on the impact velocity and angle.

The structure of this paper is as follows: In Section 2, we introduce SPH method and models for realistic rocky material. In Section 3, we describe initial conditions of impact simulations and the way to analyze results. Detailed results are introduced in Section 4, and in Section 5 we discuss physical explanation for our results and application. In section 6, we summarize our findings. 

\section{Method}

\subsection{SPH method}
To investigate planetesimal collisions, we use SPH method for elastic dynamics (\citealt{Libersky-and-Petschek1991}). SPH method is a computational fluid dynamics method utilizing Lagrangian particles (\citealt{Monaghan1992}). In a framework of SPH method, we represent continuum material such as rock using a cluster of particles. Motion of each particle is described by the equation of motion. Each particle has physical quantities such as density and internal energy, and these physical quantities are calculated from time evolution equations such as the equation of energy.

In order to treat elastic bodies by the SPH method we use following forms of basic equations\footnote{These SPH equations are based on those of \citet{Libersky-and-Petschek1991}. In \citet{Sugiura-and-Inutsuka2016} and \citet{Sugiura-and-Inutsuka2017}, we extended the Godunov SPH method (\citealt{Inutsuka2002}) to elastic dynamics, which can suppress the tensile instability (numerical instability that is prominent in the tension dominated region) and treat strong shock waves accurately. However, the tensile instability is insignificant because of fracturing of rock, and we do not treat strong shock waves because of small impact velocities compared to sound speed. Although our Godunov SPH method has excellent capability, it needs a high computational cost. Thus, to conduct many simulations with various parameters, we utilize a usual SPH method in this paper.}:

\begin{align}
&\frac{d\rho_{i}}{dt}=-\sum_{j}m_{j}\frac{\rho_{i}}{\rho_{j}}(v_{j}^{\alpha}-v_{i}^{\alpha})\frac{\partial}{\partial x_{i}^{\alpha}}W(|\bm{x}_{i}-\bm{x}_{j}|,h), \label{EoC} \\ 
&\frac{dv_{i}^{\alpha}}{dt}=\sum_{j}m_{j}\Bigl[ \frac{\sigma_{i}^{\alpha \beta}}{\rho_{i}^{2}} + \frac{\sigma_{j}^{\alpha \beta}}{\rho_{j}^{2}} - \Pi_{ij}\delta^{\alpha \beta} \Bigr] \frac{\partial}{\partial x_{i}^{\beta}}W(|\bm{x}_{i}-\bm{x}_{j}|,h) + \sum_{j}g_{ij}^{\alpha}, \label{EoM} \\
&\frac{du_{i}}{dt}= - \sum_{j}\frac{1}{2}m_{j}\Bigl[ \frac{p_{i}}{\rho_{i}^{2}} + \frac{p_{j}}{\rho_{j}^{2}} + \Pi_{ij} \Bigl](v_{j}^{\alpha}-v_{i}^{\alpha})\frac{\partial}{\partial x_{i}^{\alpha}}W(|\bm{x}_{i}-\bm{x}_{j}|,h) \nonumber \\ & \ \ \ \ \ \ \ + \sum_{j}\frac{1}{2}m_{j}\frac{S_{i}^{\alpha \beta}}{\rho_{i}\rho_{j}} \Bigl[ (v_{j}^{\alpha}-v_{i}^{\alpha})\frac{\partial}{\partial x_{i}^{\beta}} + (v_{j}^{\beta}-v_{i}^{\beta})\frac{\partial}{\partial x_{i}^{\alpha}} \Bigr] W(|\bm{x}_{i}-\bm{x}_{j}|,h). \label{EoE} 
\end{align}

\noindent Here, $m_{i}$ is mass of the $i$-th SPH particle, $\rho_{i}$ is its density, $\bm{v}_{i}$ is its velocity vector, $\bm{x}_{i}$ is its position vector, $h$ is a smoothing length, $\sigma^{\alpha \beta}_{i}$ is its stress tensor, $u_{i}$ is its specific internal energy, $p_{i}$ is its pressure, $S^{\alpha \beta}_{i}$ is its deviatoric stress tensor, $\delta^{\alpha \beta}$ is the Kronecker delta, $W(r,h)$ is a kernel function, $\Pi_{ij}$ is artificial viscosity, and $\bm{g}_{ij}$ is the gravity between the $i$-th and $j$-th particles. Superscripts in Greek letters mean a direction or component of a vector or tensor, and subscripts in Roman letters mean the particle number. We apply the summation rule over repeated indices in Greek letters. Using the pressure $p_{i}$ and the deviatoric stress tensor $S^{\alpha \beta}_{i}$, the stress tensor $\sigma^{\alpha \beta}_{i}$ is represented as

\begin{equation}
\sigma^{\alpha \beta}_{i}=-p_{i}\delta^{\alpha\beta}+S^{\alpha\beta}_{i}.
\label{stress-tensor}
\end{equation}

\noindent For the kernel function, we use a Gaussian kernel given by

\begin{equation}
W(r,h)=\Bigl[ \frac{1}{h\sqrt{\pi}} \Bigr]^{3}\exp \Bigl( -\frac{r^{2}}{h^{2}} \Bigr).
\label{kernel-function}
\end{equation}

\noindent  We set the smoothing length to be constant because of insignificant density variation. The smoothing length is determined by initial average particle spacing.  The artificial viscosity is represented as

\begin{align}
  &\Pi_{ij} = \left\{
    \begin{array}{ll}
      \frac{-\alpha_{{\rm vis}}\mu_{ij}(C_{s,i}+C_{s,j})/2 + \beta_{{\rm vis}} \mu_{ij}^{2}}{(\rho_{i}+\rho_{j})/2} & (\bm{v}_{i}-\bm{v}_{j})\cdot (\bm{x}_{i}-\bm{x}_{j})<0\\
      0 & (\bm{v}_{i}-\bm{v}_{j})\cdot (\bm{x}_{i}-\bm{x}_{j})>0
    \end{array}
  \right. , \nonumber \\ 
  &\mu_{ij}=\frac{h(\bm{v}_{i}-\bm{v}_{j})\cdot (\bm{x}_{i}-\bm{x}_{j})}{(\bm{x}_{i}-\bm{x}_{j})^{2}+0.01h^{2}}. \label{artificial-viscosity}
\end{align}

\noindent Here, $\alpha_{{\rm vis}}$ and $\beta_{{\rm vis}}$ are parameters for the artificial viscosity. We adopt $\alpha_{{\rm vis}}=1.0$ and $\beta_{{\rm vis}}=2.0$. According to the kernel function, the gravity between the $i$-th and $j$-th particles is calculated as

\begin{align}
&\bm{g}_{ij}=-G\hat{m}_{j}\frac{\bm{x}_{i}-\bm{x}_{j}}{|\bm{x}_{i}-\bm{x}_{j}|^{3}}, \nonumber \\ 
&\hat{m}_{j}=\int_{0}^{|\bm{x}_{i}-\bm{x}_{j}|}4\pi r^{2}m_{j}W(r,h)dr, \label{self-gravity}
\end{align}

\noindent  where $G$ is the gravitational constant.  According to the Hooke's law, deviatoric stress is proportional to strain. We calculate the time evolution of strain and then obtain the time evolution of deviatoric stress. The time evolution equation for the deviatoric stress tensor is given by

\begin{equation}
\frac{dS_{i}^{\alpha \beta}}{dt}=2\mu \Bigl( \epsilon_{i}^{\alpha \beta} - \frac{1}{3}\epsilon_{i}^{\gamma \gamma}\delta^{\alpha \beta} \Bigr) + S_{i}^{\alpha \gamma}R_{i}^{\beta \gamma} + S_{i}^{\beta \gamma}R_{i}^{\alpha \gamma},
\label{Hooke-law}
\end{equation}

\noindent where $\mu$ is the shear modulus. $\epsilon_{i}^{\alpha \beta}$ and $R_{i}^{\alpha \beta}$ are a strain rate tensor and a rotational rate tensor respectively, and represented as

\begin{align}
& \epsilon_{i}^{\alpha \beta}=\frac{1}{2}\Bigl( \frac{\partial v_{i}^{\alpha}}{\partial x_{i}^{\beta}} + \frac{\partial v_{i}^{\beta}}{\partial x_{i}^{\alpha}} \Bigr), \label{strain-rate-tensor} \\ 
& R_{i}^{\alpha \beta}=\frac{1}{2}\Bigl( \frac{\partial v_{i}^{\alpha}}{\partial x_{i}^{\beta}} - \frac{\partial v_{i}^{\beta}}{\partial x_{i}^{\alpha}} \Bigr). \label{rotational-rate-tensor}
\end{align}

\noindent Note that $\epsilon_{i}^{\alpha \beta}$ and $R_{i}^{\alpha \beta}$ are described by sums of velocity gradients. To treat rigid body rotation correctly, we adopt equations of velocity gradients with the correction matrix $\bm{{\rm L}}_{i}$ developed by \citet{Bonet-and-Lok1999}:

\begin{align}
& \frac{\partial v_{i}^{\alpha}}{\partial x_{i}^{\beta}} = \sum_{j}\frac{m_{j}}{\rho_{j}}(v_{j}^{\alpha}-v_{i}^{\alpha}){\rm L}_{i}^{\beta \gamma}\frac{\partial}{\partial x_{i}^{\gamma}}W(|\bm{x}_{i}-\bm{x}_{j}|,h), \nonumber \\ 
& \bm{{\rm L}}_{i}=\Bigl( \sum_{j}\frac{m_{j}}{\rho_{j}}\frac{\partial}{\partial \bm{x}_{i}}W(|\bm{x}_{i}-\bm{x}_{j}|,h) \otimes (\bm{x}_{j}-\bm{x}_{i}) \Bigr)^{-1}. \label{corrected-velocity-gradient}
\end{align}

For the equation of state, we use the Tillotson EoS (\citealt{Tillotson1962}), which is often used for numerical simulations of impacts with SPH method (e.g., \citealt{Genda-et-al2012, Benz-and-Asphaug1999}). The Tillotson EoS has ten material dependent parameters. We assume a material of colliding planetesimals as basalt. Thus we use values of the shear modulus and the Tillotson parameters for basalt described in \citet{Benz-and-Asphaug1999}.

For time integration, we use a leapfrog method with a kick-drift-kick scheme. Here, we use leapfrog equations with a form where the position and other physical quantities such as the velocity are both updated at the end of each step (e.g., \citealt{Hubber-et-al2013}). Detailed description for the time integration scheme is given in Appendix A.

For fast calculation of the time evolution equations (\ref{EoC}), (\ref{EoM}) and (\ref{EoE}), we parallelize our simulation code using Framework for Developing Particle Simulator (FDPS; \citealt{Iwasawa-et-al2015, Iwasawa-et-al2016}). FDPS is a framework to support developing efficiently-parallelized simulation codes that are based on particle methods. FDPS provides functions for exchanging information of particles between CPUs and those for load balancing. Therefore, thanks to FDPS, we effectively use parallel computers for our simulations.

\subsection{Models for the fracture and friction}
To treat the collisional destruction of rocky material, we apply appropriate models for the fracture of rock and the friction between completely damaged material.

\citet{Benz-and-Asphaug1995} introduce a fracture model based on the model for brittle solid (\citealt{Grady-and-Kipp1980}) to SPH method. According to this model, we use a damage parameter $D$. Each SPH particle has this state variable $D$. SPH particles with $D=0$ represent intact rock, and those with $D=1$ represent completely damaged rock, which means that these SPH particles do not feel any tensile stress.  The damage parameter increases according to the function modeled by \citet{Benz-and-Asphaug1995} if local strain exceeds flaw's activation threshold.  Flaw's activation threshold is determined by material dependent parameters and total volume of rock. For these parameters we also use values for basalt described in \citet{Benz-and-Asphaug1999}.

 According to the fracture model, we modify the pressure and use damage relieved pressure $p_{d,i}$;

\begin{equation}
  p_{d,i} = \left\{
    \begin{array}{ll}
      (1-D)p_{i} & p_{i}<0\\
      p_{i} & p_{i}>0
    \end{array}
  \right. 
\label{relieved-pressure}
\end{equation}

\noindent for Eqs.\,(\ref{EoM}) and (\ref{EoE}). 

 We treat the friction of damaged rock ($D>0$) according to \citet{Jutzi2015}. For collisions of our interest, the energy dissipation by the friction of partially damaged rock ($0<D<1$) is much smaller than that of completely damaged rock ($D=1$). Therefore, we only explain the treatment for the friction of completely damaged rock. 

To represent the friction of granular materials, we set yielding strength $Y_{d,i}$ as

\begin{equation}
  Y_{d,i}=\mu_{d}p_{d,i},
  \label{yield-strength-for-D=1-particle}
\end{equation}

\noindent where $\mu_{d}$ is the friction coefficient. Here we assume $\mu_{d}=\tan (40\degr) = 0.839$, which corresponds to a material with the angle of friction of $40\degr$. Note that the angle of friction of lunar sand is estimated to be $30\degr - 50\degr$ (e.g., \citealt{Heiken-et-al1991}), which is consistent with the angle of friction estimated from surface slopes of asteroid Itokawa (\citealt{Fujiwara-et-al2006}). Using the yielding strength of Eq.\,(\ref{yield-strength-for-D=1-particle}), we modify the deviatoric stress tensor as

\begin{align}
& S^{\alpha \beta}_{i}\rightarrow f_{i}S^{\alpha \beta}_{i}, \nonumber \\ 
& f_{i}=\min [Y_{d,i}/\sqrt{J_{2,i}},1], \nonumber \\
& J_{2,i}=\frac{1}{2}S^{\alpha\beta}_{i}S^{\alpha\beta}_{i}. \label{friction-model}
\end{align}

\noindent Owing to this friction model, the formation of irregular shapes of rubble piles are reproduced.

\section{Initial conditions of impacts and analysis of results}

\subsection{Initial conditions of impacts}
For simplicity, we use a sphere of basalt with zero rotation as an initial planetesimal. The radius of planetesimals is set to $R_{t}=50\,{\rm km}$, and we focus on collisions between two equal-mass planetesimals with mass of $M_{{\rm target}}=4\pi \rho_{0} R_{t}^{3}/3$, where $\rho_{0}$ is the uncompressed density of basalt. We carry out relatively low-resolution simulations using 100,000 SPH particles for a simulation, which enables to reveal detailed dependence of resultant shapes on impact velocities and angles. Validity of this number of SPH particles is discussed in Section 4.1.

For a basaltic planetesimal with the radius of $50\,{\rm km}$, the density at the center in hydrostatic equilibrium is almost the same as uncompressed density. Thus we set initial planetesimals to be uniform spheres with the mean density of basalt. To do so, an isotropic SPH particle distribution is more preferable than, for example, particles placed on cubic lattices, so that we prepare a particle distribution with uniform disposition from a random distribution. Detailed procedures to produce the uniform and isotropic particle distribution are as follows: Firstly we put SPH particles within a cubic domain with periodic boundary conditions randomly so that desired resolution and desired mean density are achieved. Secondly we let the particles move under the forces anti-parallel to density gradients that make the particle distribution uniform until the standard deviation of density becomes less than 0.1\% of the mean density. Finally, we remove particles outside a shell with the radius of $50\,{\rm km}$, and then a uniform and isotropic sphere is obtained.

\begin{figure}[!htb]
  \begin{center}
    \includegraphics[bb=0 0 323 285, width=0.6\linewidth,clip]{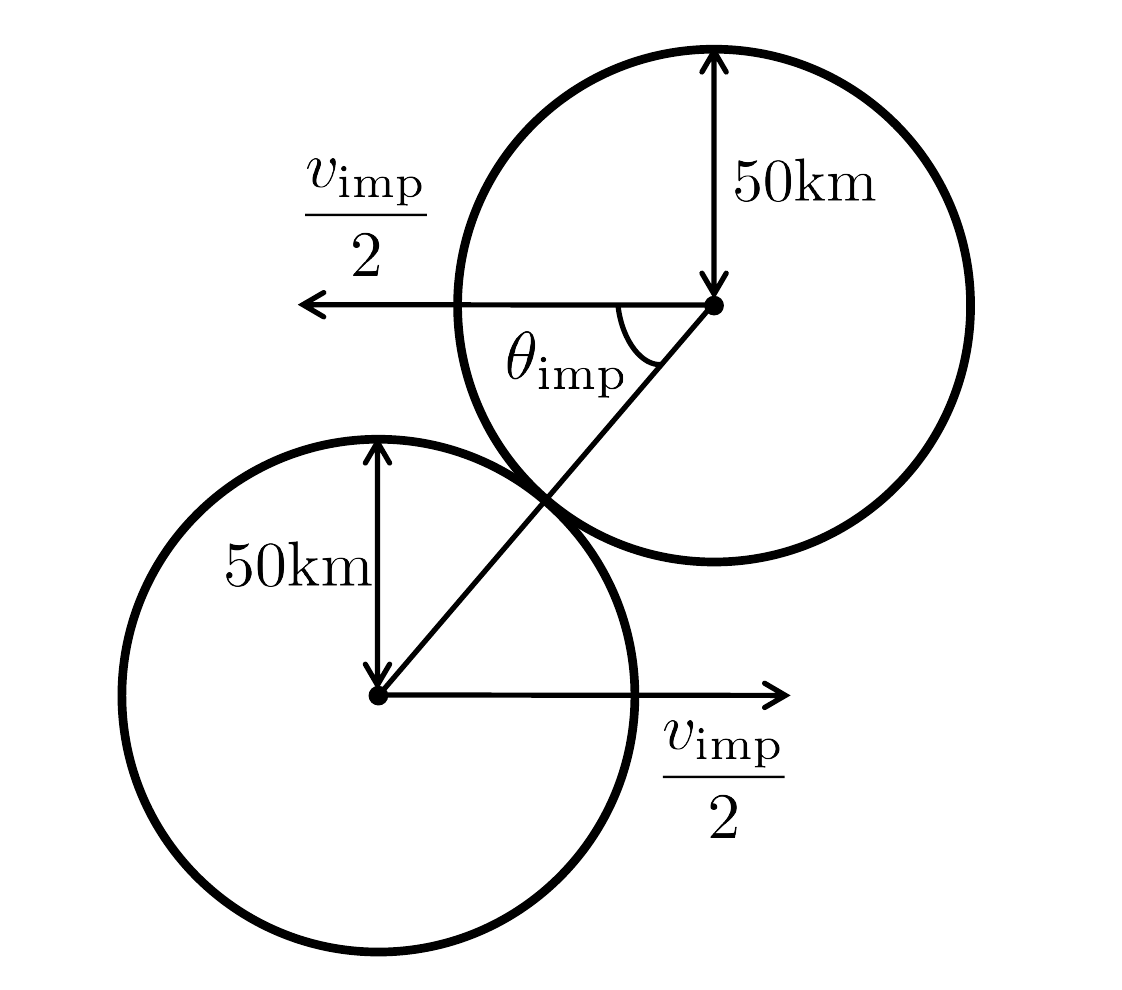}
    \caption{Impact geometry and the definition of the impact velocity $v_{{\rm imp}}$ and angle $\theta_{{\rm imp}}$.}
    \label{Schematic-of-impact-condition}
  \end{center}
\end{figure}

We define the impact velocity $v_{{\rm imp}}$ as the relative velocity between two planetesimals at the time of impact, and the impact angle $\theta_{{\rm imp}}$ as the angle between the line joining centers of two planetesimals and the relative velocity vector at the time of impact. Thus the impact angle of $0\degr$ means a head-on collision, and that of $90\degr$ means a grazing collision. Fig.\,\ref{Schematic-of-impact-condition} schematically shows the definition of the impact velocity and angle. At the beginning of simulations, centers of two planetesimals are apart at a distance of $4R_{t}$.

\subsection{Analysis of results}
We conduct simulations of impacts and subsequent gravitational reaccumulation over a period of $1.0\times 10^{5}\,{\rm s}$. The typical timescale of reaccumulation is estimated as $t_{{\rm acc}}=2R_{t}/v_{{\rm esc}}$, where $v_{{\rm esc}}$ is the two-body escape velocity of planetesimals. The value of $t_{{\rm acc}}$ is calculated as

\begin{align}
  t_{{\rm acc}}=\frac{2R_{t}}{v_{{\rm esc}}}=\sqrt{\frac{3}{2\pi G \rho_{0}}} \simeq 1600\,{\rm s}. \label{reaccumulation-timescale}
\end{align}

\noindent Thus $1.0\times 10^{5}\,{\rm s}$ is about 100 times longer than the typical timescale of reaccumulation, and we also confirmed that gravitational reaccumulation is sufficiently finished after $1.0\times 10^{5}\,{\rm s}$.

After collisional simulations, we identify the largest remnants using a friends-of-friends algorithm (e.g., \citealt{Huchra-and-Geller1982}). We find a swarm of SPH particles with spacing less than $1.5h$ and then the largest swarm is identified with the largest remnant. 

Then we evaluate the shapes of the largest remnants. To do so, we quantitatively measure the axis lengths of the largest remnants using the inertia moment tensor. We approximate the largest remnant as an ellipsoid that has the same inertia moment tensor and mass, and then we identify the axis lengths of the ellipsoid with those of the largest remnant. Detailed procedure to calculate axis lengths is given in Appendix B. It should be noted that the bodies resulting from simulations are not perfect ellipsoids. The obtained axis ratios are thus different from those measured in the top-down method that is usually used in laboratory experiments. Therefore the axis ratios include measurement errors of $\sim 0.1$ (see \citealt{Michikami-et-al2018}).

Shapes of objects are characterized by ratios between the lengths of major axis $a$, intermediate axis $b$, and minor axis $c$, i.e., $b/a$ and $c/a$ ($b/a$ and $c/a$ are $0 - 1$ and $b/a > c/a$ by definition). Bodies with $c/a \sim 1$ have almost spherical shapes. For $b/a \sim 1$ and $c/a \ll 1$, bodies have flat shapes. Bodies with $b/a \ll 1$ have elongated shapes. 

\section{Results}

\subsection{Resolution dependence on the resultant shape}
\begin{figure}[!htb]
  \begin{center}
    \includegraphics[bb=0 0 960 498, width=1.0\linewidth,clip]{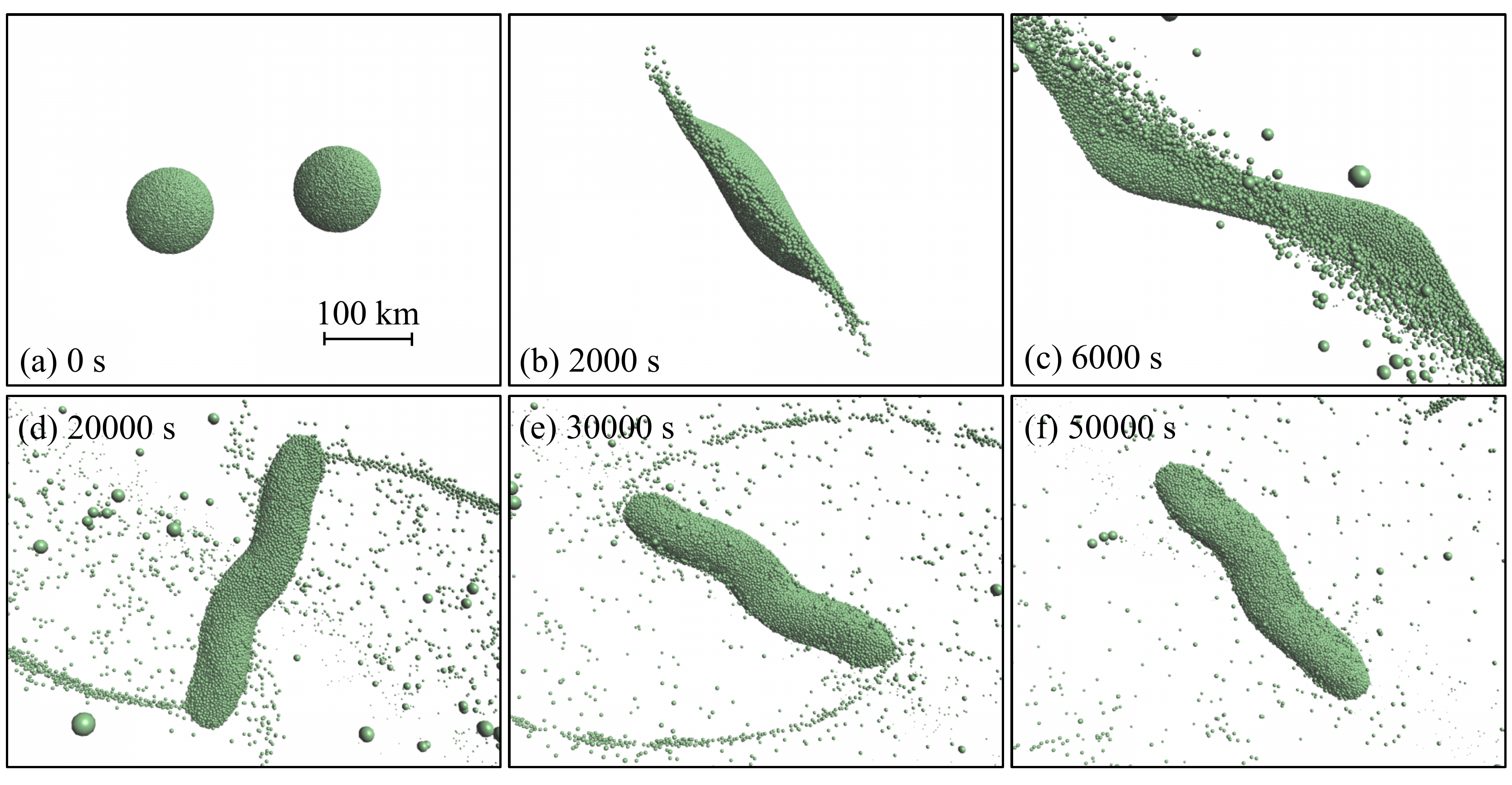}
    \caption{Snapshots of the impact simulation with the impact angle $\theta_{{\rm imp}}$ of $15\degr$, the impact velocity $v_{{\rm imp}}$ of $200\,{\rm m/s}$ and the total number of SPH particles $N_{{\rm total}}$ of $1\times 10^{5}$ at $0.0\,{\rm s}$(a), $2.0\times 10^{3}\,{\rm s}$(b), $6.0\times 10^{3}\,{\rm s}$(c), $2.0\times 10^{4}\,{\rm s}$(d), $3.0\times 10^{4}\,{\rm s}$(e), and $5.0\times 10^{4}\,{\rm s}$(f). Scale on Panel (a) is also valid for all Panels (b)-(f).}
    \label{subsequent-pictures-50-1.0-15-200-100000}
  \end{center}
\end{figure}

Figure \ref{subsequent-pictures-50-1.0-15-200-100000} represents snapshots of the SPH simulation with the impact angle $\theta_{{\rm imp}}$ of $15\degr$ and the impact velocity $v_{{\rm imp}}$ of $200\,{\rm m/s}$. In Fig.\,\ref{subsequent-pictures-50-1.0-15-200-100000}b, the collision induces shattering of planetesimals. Then two planetesimals are stretched in the direction perpendicular to the line joining centers of two contacting planetesimals and fragments are ejected (Fig.\,\ref{subsequent-pictures-50-1.0-15-200-100000}c). Ejected materials are mainly reaccumulated from the direction of the long axis of the largest remnant (Fig.\,\ref{subsequent-pictures-50-1.0-15-200-100000}d). Finally a very elongated shape with the ratio $b/a$ of about 0.2 is formed (Fig.\,\ref{subsequent-pictures-50-1.0-15-200-100000}f).  The accretion on the largest body is mostly done within $t\sim 5.0\times 10^{4}\,{\rm s}$. 

\begin{figure}[!htb]
  \begin{center}
    \includegraphics[bb=0 0 960 257, width=1.0\linewidth,clip]{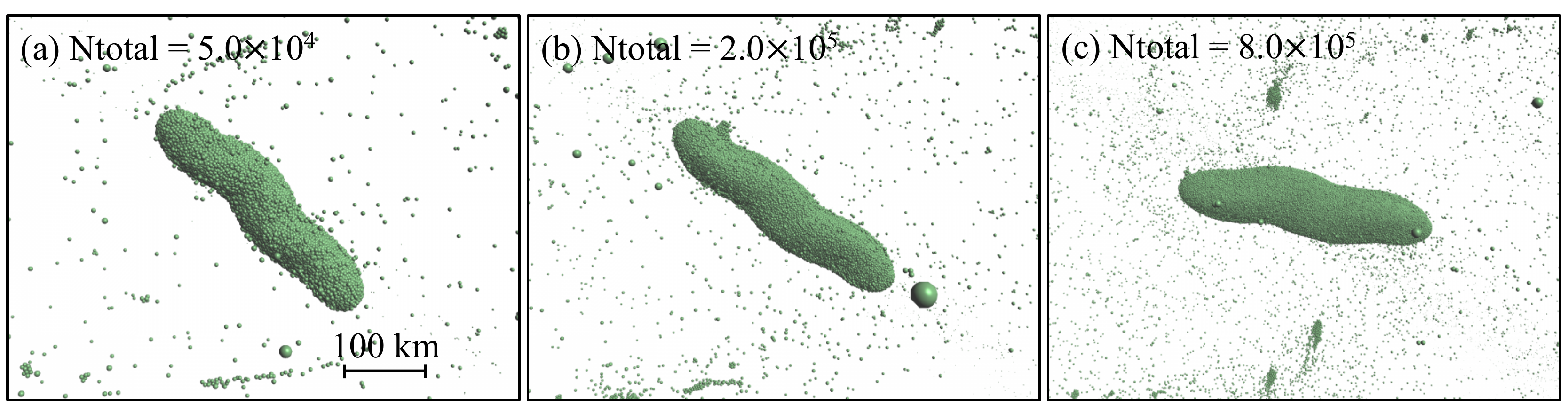}
    \caption{Shapes of the largest remnants at $5.0\times 10^{4}\,{\rm s}$ for the impact simulations with $v_{{\rm imp}}=200\,{\rm m/s}$, $\theta_{{\rm imp}}=15\degr$, and $N_{{\rm total}}=5\times 10^{4}$(a), $2\times 10^{5}$(b), and $8\times 10^{5}$(c), respectively.}
    \label{pictures-50-1.0-15-200-resolution-dependence}
  \end{center}
\end{figure}

\begin{figure}[!htb]
 \begin{center}
 \includegraphics[bb=0 0 792 540, width=1.0\linewidth,clip]{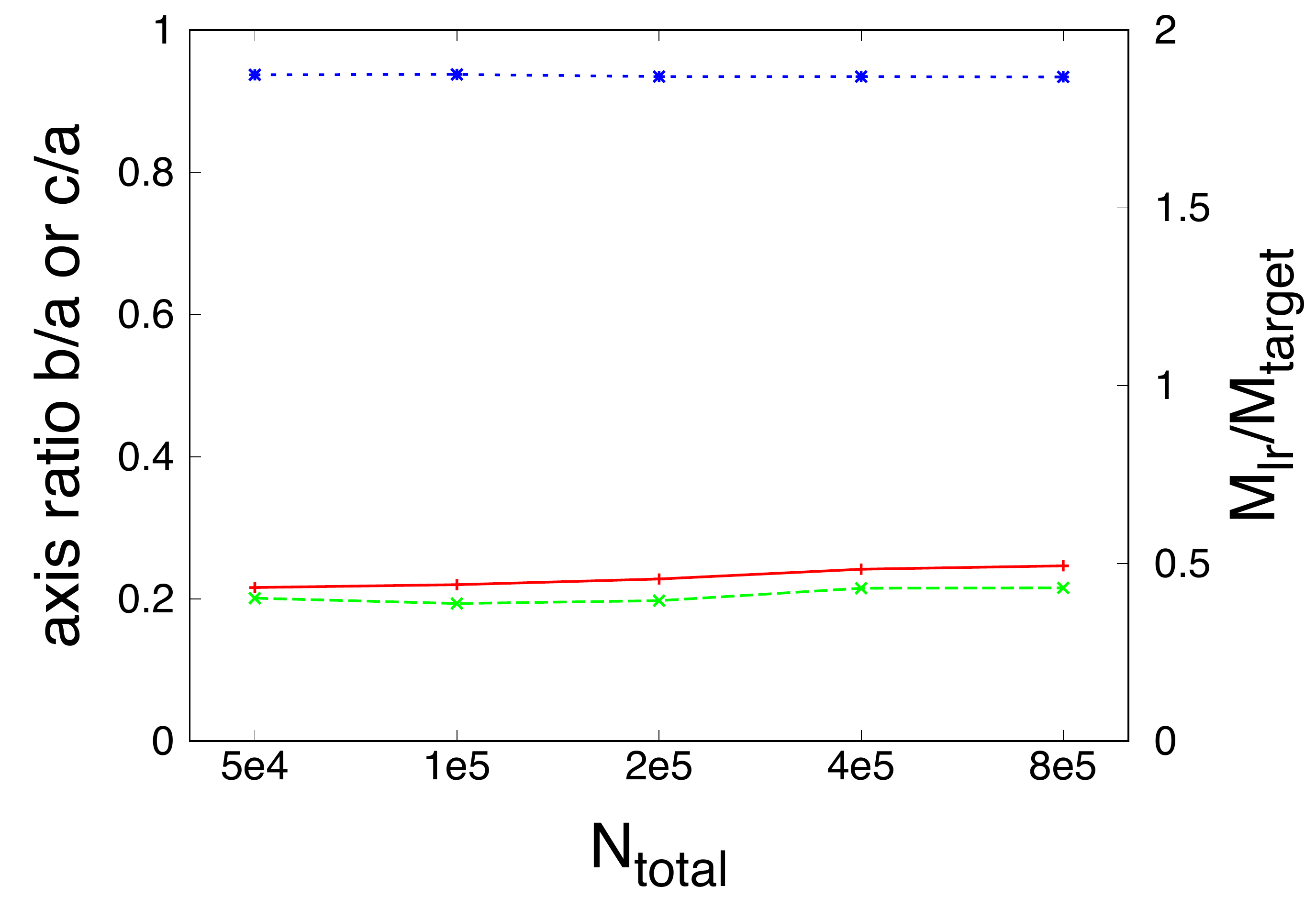}
 \caption{Dependence of the mass and axis ratios of the largest remnants on the number of SPH particles $N_{{\rm total}}$ for the impact with $\theta_{{\rm imp}} = 15\degr$ and $v_{{\rm imp}}=200\,{\rm m/s}$. Red solid line shows the ratio $b/a$, green dashed line shows the ratio $c/a$, and blue dotted line shows the mass of the largest remnants $M_{{\rm lr}}$ normalized by the mass of an initial planetesimal $M_{{\rm target}}$. Left vertical axis shows the axis ratios, and right vertical axis shows the mass of the largest remnant.}
 \label{resolution-dependence-50-1.0-15-200-cartesian}
 \end{center}
\end{figure}

Figure \ref{pictures-50-1.0-15-200-resolution-dependence} shows shapes of the largest remnants at $5.0\times 10^{4}\,{\rm s}$ with three different resolutions  (the total number of SPH particles $N_{{\rm total}}$ of $5\times 10^{4}$(a), $2\times 10^{5}$(b), and $8\times 10^{5}$(c)). Even if $N_{{\rm total}}$ becomes ten times larger, the characteristic of elongated shape does not change. Figure \ref{resolution-dependence-50-1.0-15-200-cartesian} shows the dependence of the mass and axis ratios of the largest remnants on the number of SPH particles $N_{{\rm total}}$. The mass of the largest remnants slightly decreases with increasing $N_{{\rm total}}$ because numerical dissipation by the artificial viscosity decreases for higher resolution. This tendency is the same as the result of \citet{Genda-et-al2015}. The axis ratios slightly increase with increasing $N_{{\rm total}}$, and the difference of $b/a$ between $N_{{\rm total}}$ of $5\times 10^{4}$ and $8\times 10^{5}$ is about 0.03. Difference of axis ratios less than 0.1 is unimportant for the analysis of asteroidal shapes because difference of axis measurements also causes such minor errors as discussed above. Therefore, the number of SPH particles of $10^{5}$ is sufficient to capture at least the feature of shapes. 

\subsection{Mass of the largest remnants}
Hereafter, we use $10^{5}$ SPH particles for a simulation, and we measure the mass and axis ratios of the largest remnants at $1.0\times 10^{5}\,{\rm s}$ after impacts.

\begin{figure}[!htb]
 \begin{center}
 \includegraphics[bb=0 0 764 540, width=1.0\linewidth,clip]{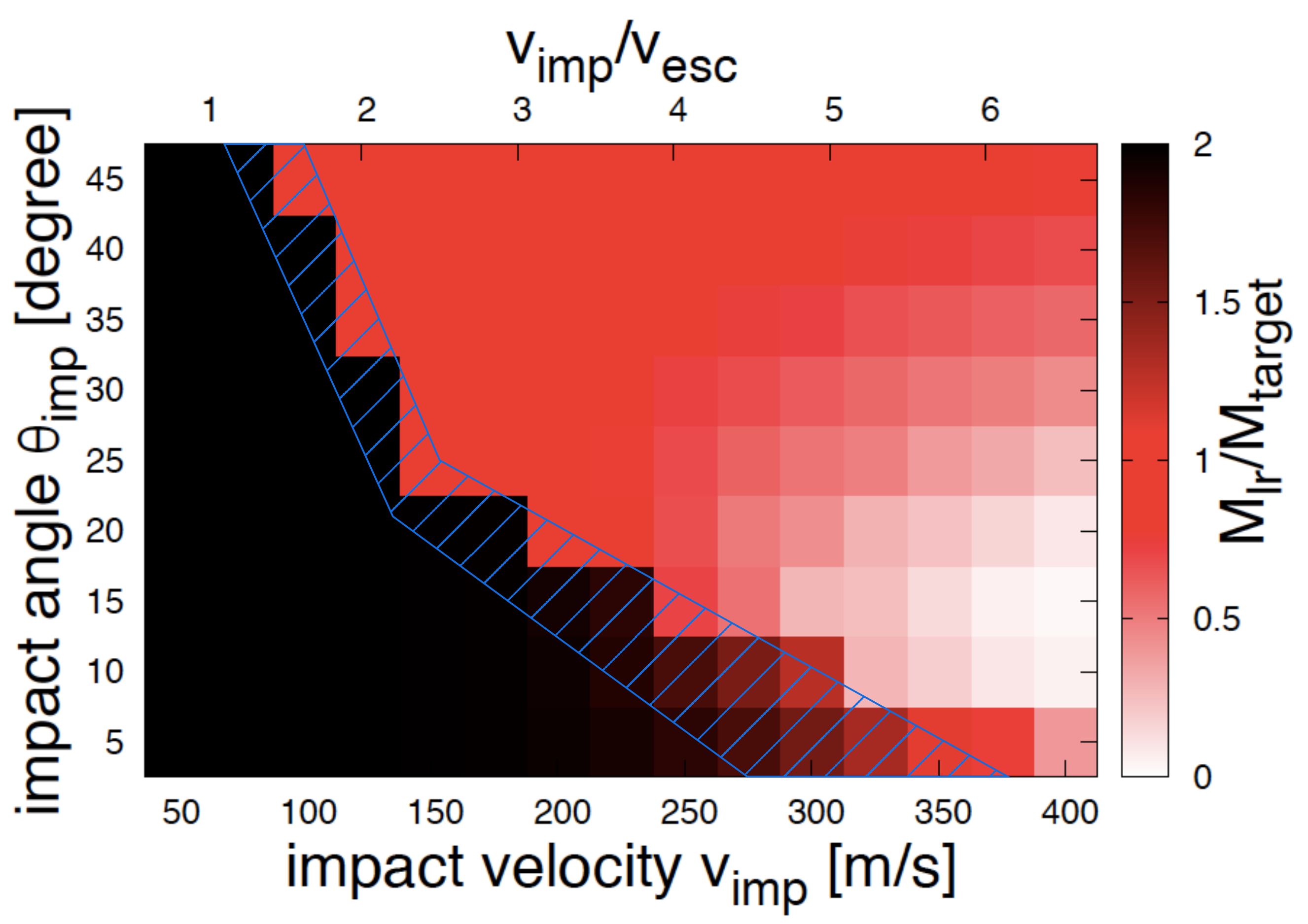}
 \caption{Dependence of the mass of the largest remnants on the impact velocity $v_{{\rm imp}}$ and the impact angle $\theta_{{\rm imp}}$. Upper horizontal axis shows $v_{{\rm imp}}$ normalized by two-body escape velocity $v_{{\rm esc}}$. Color shows the mass of the largest remnants $M_{{\rm lr}}$ normalized by the mass of an initial planetesimal $M_{{\rm target}}$. Thus $M_{{\rm lr}}/M_{{\rm target}}=2.0$ means complete merging. The hatched region apploximately shows the transitional parameters from merging collisions to hit-and-run or elosive collisions.}
 \label{fmlr-contour-q=1.0-largest-only-detailed}
 \end{center}
\end{figure}

Figure \ref{fmlr-contour-q=1.0-largest-only-detailed} shows the mass of the largest remnants $M_{{\rm lr}}$ formed through collisions with $v_{{\rm imp}} = 50\,{\rm m/s} - 400\,{\rm m/s}$ and $\theta_{{\rm imp}} = 5\degr - 45\degr$. The increment of the velocity is $25\,{\rm m/s}$, and that of the angle is $5\degr$. For $\theta_{{\rm imp}} \leq 15\degr$, $M_{{\rm lr}}\sim 2M_{{\rm target}}$ due to collisional merging for low $v_{{\rm imp}}$, and $M_{{\rm lr}}$ gradually decreases with increasing $v_{{\rm imp}}$ because of erosive collisions. The impact parameters for the transition between merging and erosive collisions are highlighted in Fig.\,\ref{fmlr-contour-q=1.0-largest-only-detailed}. On the other hand, for $\theta_{{\rm imp}} \geq 20\degr$, sharp variation in $M_{{\rm lr}}$ from $\sim 2M_{{\rm target}}$ to $\sim M_{{\rm target}}$ is seen around $v_{{\rm imp}}\sim 100\,{\rm m/s}$.  This is because collisions with high $v_{{\rm imp}}$ result in a ``hit-and-run'' process where two planetesimals moving apart after the collision. The transition parameters between merging and hit-and-run collisions are also highlighted in Fig.\,\ref{fmlr-contour-q=1.0-largest-only-detailed}. Erosive nature for low $\theta_{{\rm imp}}$ and merging/hit-and-run nature for high $\theta_{{\rm imp}}$ are also observed in previous collisional simulations (\citealt{Agnor-and-Asphaug2004,Leinhardt-and-Stewart2012}).

For $v_{{\rm imp}} > 300\,{\rm m/s}$, $M_{{\rm lr}}$ has a minimum value at $\theta_{{\rm imp}} \approx 15\degr$ (see Fig.\,\ref{fmlr-contour-q=1.0-largest-only-detailed}). For head-on collisions, the most of the impact energy is dissipated and not transformed to the ejection processes, which results in large $M_{{\rm lr}}$. For slightly higher $\theta_{{\rm imp}}$, the impact energy is more effectively used for ejection, and thus the mass of the largest remnant $M_{{\rm lr}}$ becomes smaller. However, for much  higher $\theta_{{\rm imp}}$, the velocity component normal to colliding bodies is small, so that the impact energy is not effectively used for destruction and ejection, which results in larger $M_{{\rm lr}}$. Therefore, an intermediate $\theta_{{\rm imp}}$ yields smallest $M_{{\rm lr}}$.

We note that collisions with $v_{{\rm imp}}>400\,{\rm m/s}$ and low $\theta_{{\rm imp}}$ result in $M_{{\rm lr}} \leq 0.1M_{{\rm target}}$. The largest remnants resulting from such impacts are composed of less than about 5,000 SPH particles, and resolved by less than 20 SPH particles along each axis direction. Thus axis ratios obtained from such a small number of SPH particles are not measured accurately. Our simulations of impacts with parameters outside those of Fig.\ref{fmlr-contour-q=1.0-largest-only-detailed} show that impacts with $v_{{\rm imp}}=500\,{\rm m/s}$ and $\theta_{{\rm imp}}= 5 - 25\degr$ result in $M_{{\rm lr}} / M_{{\rm target}} = 0.01 - 0.07$. For $\theta_{{\rm imp}} > 45\degr$, only edges of planetesimals are destroyed by collisions rather than overall deformation, so that the investigation of such impact angles is not interesting. For example, our impact simulations with $\theta_{{\rm imp}}=60\degr$ and $v_{{\rm imp}}=100 - 500\,{\rm m/s}$ result in $M_{{\rm lr}} / M_{{\rm target}} = 0.91 - 0.99$, which means merely partial destruction. Therefore, we investigate the collisions with $50\,{\rm m/s}\leq v_{{\rm imp}} \leq 400\,{\rm m/s}$ and $5\degr\leq \theta_{{\rm imp}} \leq 45\degr$, because in this parameter range the resolution of the largest remnants is mainly sufficient and significant shape deformation occurs.

\subsection{Characteristic shapes formed by collisions}
As a result of impact simulations with $50\,{\rm m/s}\leq v_{{\rm imp}} \leq 400\,{\rm m/s}$ and $5\degr\leq \theta_{{\rm imp}} \leq 45\degr$, we find that resultant shapes of the largest remnants are roughly classified into five categories. In this subsection, we introduce the results of typical impacts to form five different characteristic shapes and catastrophic collisions.

\subsubsection{Bilobed shapes}
\begin{figure}[!htb]
  \begin{center}
    \includegraphics[bb=0 0 648 498, width=0.8\linewidth,clip]{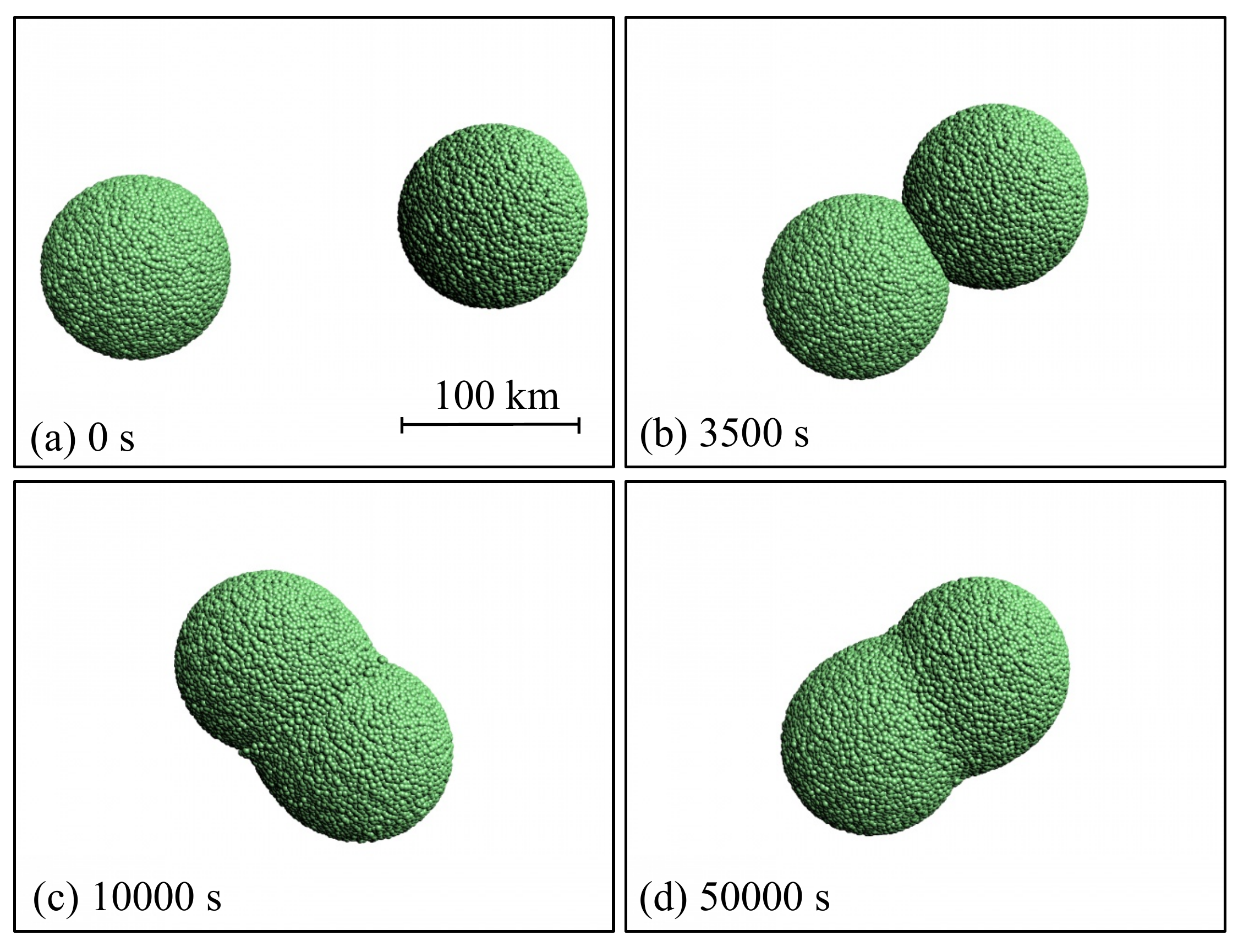}
    \caption{Snapshots of the impact simulation with $\theta_{{\rm imp}}=30\degr$ and $v_{{\rm imp}}=50\,{\rm m/s}$ at $0.0\,{\rm s}$(a), $3.5\times 10^{3}\,{\rm s}$(b), $1.0\times 10^{4}\,{\rm s}$(c), and $5.0\times 10^{4}\,{\rm s}$(d).}
    \label{subsequent-pictures-50-1.0-30-50}
  \end{center}
\end{figure}

If the impact velocity is very small, the initial spherical shapes of colliding bodies are preserved and collisional merging forms bilobed shape. Fig.\,\ref{subsequent-pictures-50-1.0-30-50} shows impact snapshots with $v_{{\rm imp}} = 50\,{\rm m/s}$ and $\theta_{{\rm imp}} = 30\degr$. The impact forms a bilobed shape (Fig.\,\ref{subsequent-pictures-50-1.0-30-50}). The two-body escape velocity $v_{{\rm esc}}$ is about $60\,{\rm m/s}$, which is slightly larger than the impact velocity of this simulation. For $v_{{\rm imp}} < v_{{\rm esc}}$, the impact energy is too small to largely deform the initial spherical shapes (see Fig.\,\ref{subsequent-pictures-50-1.0-30-50}b,c), and colliding bodies are gravitationally bound. Thus the bilobed shapes resulting from such low velocity impacts are independent of $\theta_{{\rm imp}}$.

\subsubsection{Spherical shapes}
\begin{figure}[!htb]
  \begin{center}
    \includegraphics[bb=0 0 705 540, width=0.8\linewidth,clip]{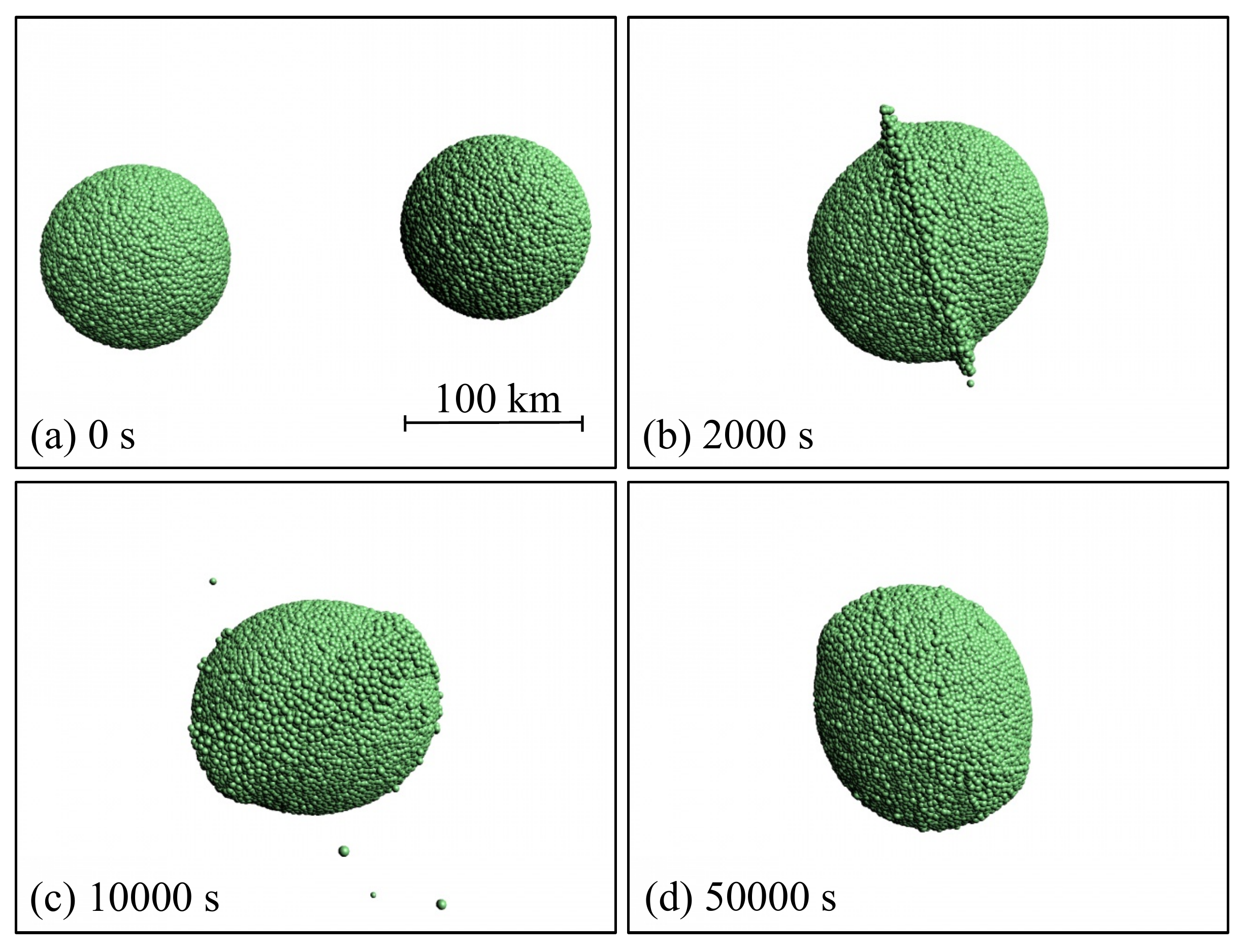}
    \caption{Snapshots of the impact simulation with $\theta_{{\rm imp}}=10\degr$ and $v_{{\rm imp}}=100\,{\rm m/s}$ at $0.0\,{\rm s}$(a), $2.0\times 10^{3}\,{\rm s}$(b), $1.0\times 10^{4}\,{\rm s}$(c), and $5.0\times 10^{4}\,{\rm s}$(d).}
    \label{subsequent-pictures-50-1.0-10-100}
  \end{center}
\end{figure}

The initial spherical shape is sufficiently deformed with $v_{{\rm imp}}\sim 100\,{\rm m/s}$, which results in a single sphere due to merging of two planetesimals. Fig.\,\ref{subsequent-pictures-50-1.0-10-100} shows an impact producing a spherical shape with $v_{{\rm imp}}=100\,{\rm m/s}$ and $\theta_{{\rm imp}}=10\degr$. Collisional deformation (Fig.\,\ref{subsequent-pictures-50-1.0-10-100}b) and gravitational reaccumulation (Fig.\,\ref{subsequent-pictures-50-1.0-10-100}c,d) results in a spherical shape.

It should be noted that a relatively low speed collision with $\theta_{{\rm imp}} \geq 40\degr$ results in local destruction due to hit-and-run, whose outcome is also close to two spheres.

\subsubsection{Flat shapes}

\begin{figure}[!htb]
  \begin{center}
    \includegraphics[bb=0 0 960 498, width=1.0\linewidth,clip]{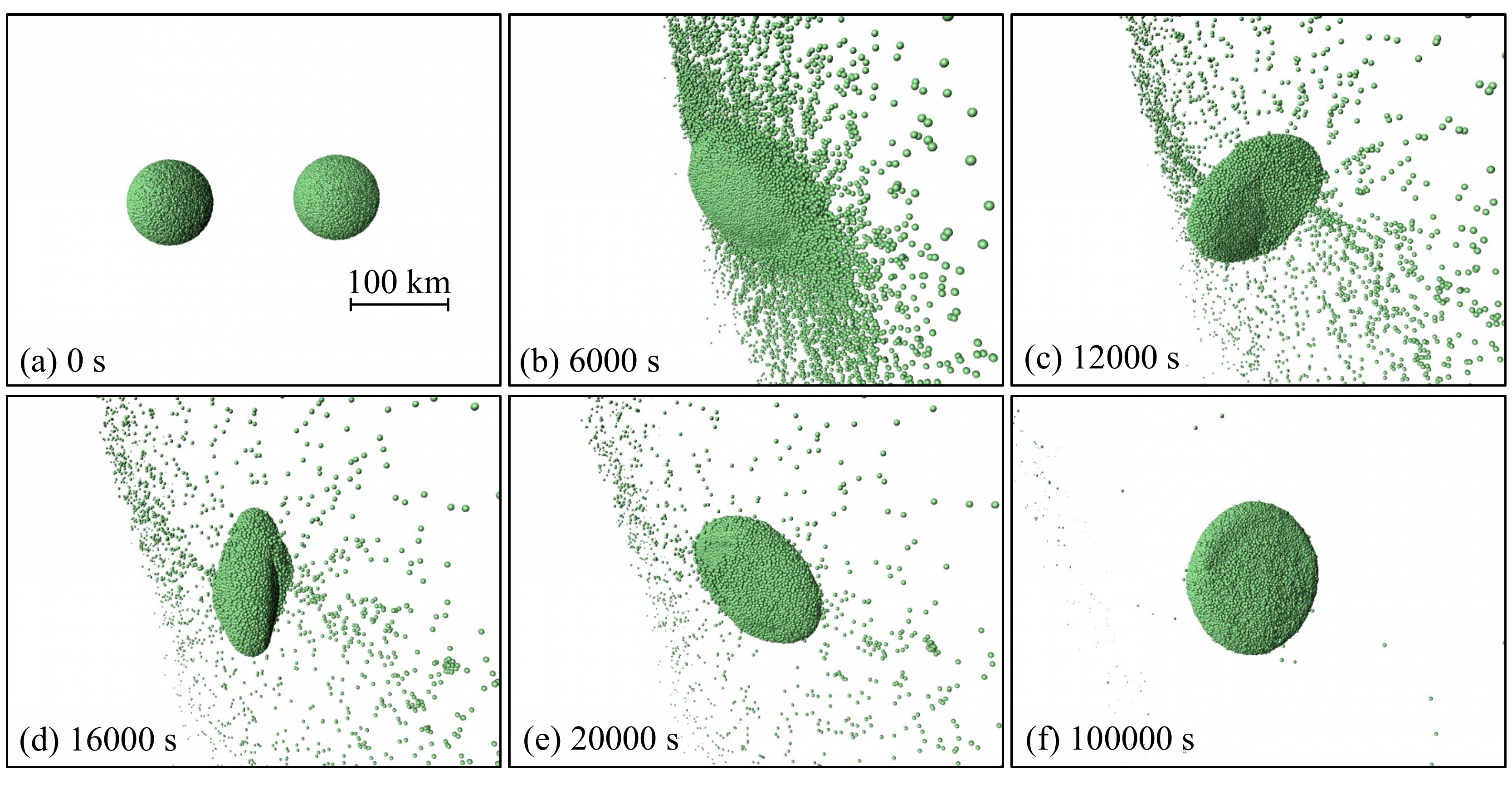}
    \caption{Snapshots of the impact simulation with $\theta_{{\rm imp}}=5\degr$ and $v_{{\rm imp}}=200\,{\rm m/s}$ at $0.0\,{\rm s}$(a), $6.0\times 10^{3}\,{\rm s}$(b), $1.2\times 10^{4}\,{\rm s}$(c), $1.6\times 10^{4}\,{\rm s}$(d), $2.0\times 10^{4}\,{\rm s}$(e), and $1.0\times 10^{5}\,{\rm s}$(f).}
    \label{subsequent-pictures-50-1.0-5-200}
  \end{center}
\end{figure}

Figure \ref{subsequent-pictures-50-1.0-5-200} shows impact snapshots with $v_{{\rm imp}}=200\,{\rm m/s}$ and $\theta_{{\rm imp}}=5\degr$. The initial spherical shapes are completely deformed (Fig.\,\ref{subsequent-pictures-50-1.0-5-200}b,c), and the resultant shape is flat (Fig.\,\ref{subsequent-pictures-50-1.0-5-200}d-f). The flat bodies are close to oblate shapes. The minor axis is formed in the direction perpendicular to the angular momentum vector. 

\subsubsection{Elongated shapes}
A collision forming extremely elongated shape is shown in Fig.\,\ref{subsequent-pictures-50-1.0-15-200-100000}. The collision results in $M_{{\rm lr}} \sim 2M_{{\rm target}}$; collisional merging mainly occurs.

\begin{figure}[!htb]
  \begin{center}
    \includegraphics[bb=0 0 960 498, width=1.0\linewidth,clip]{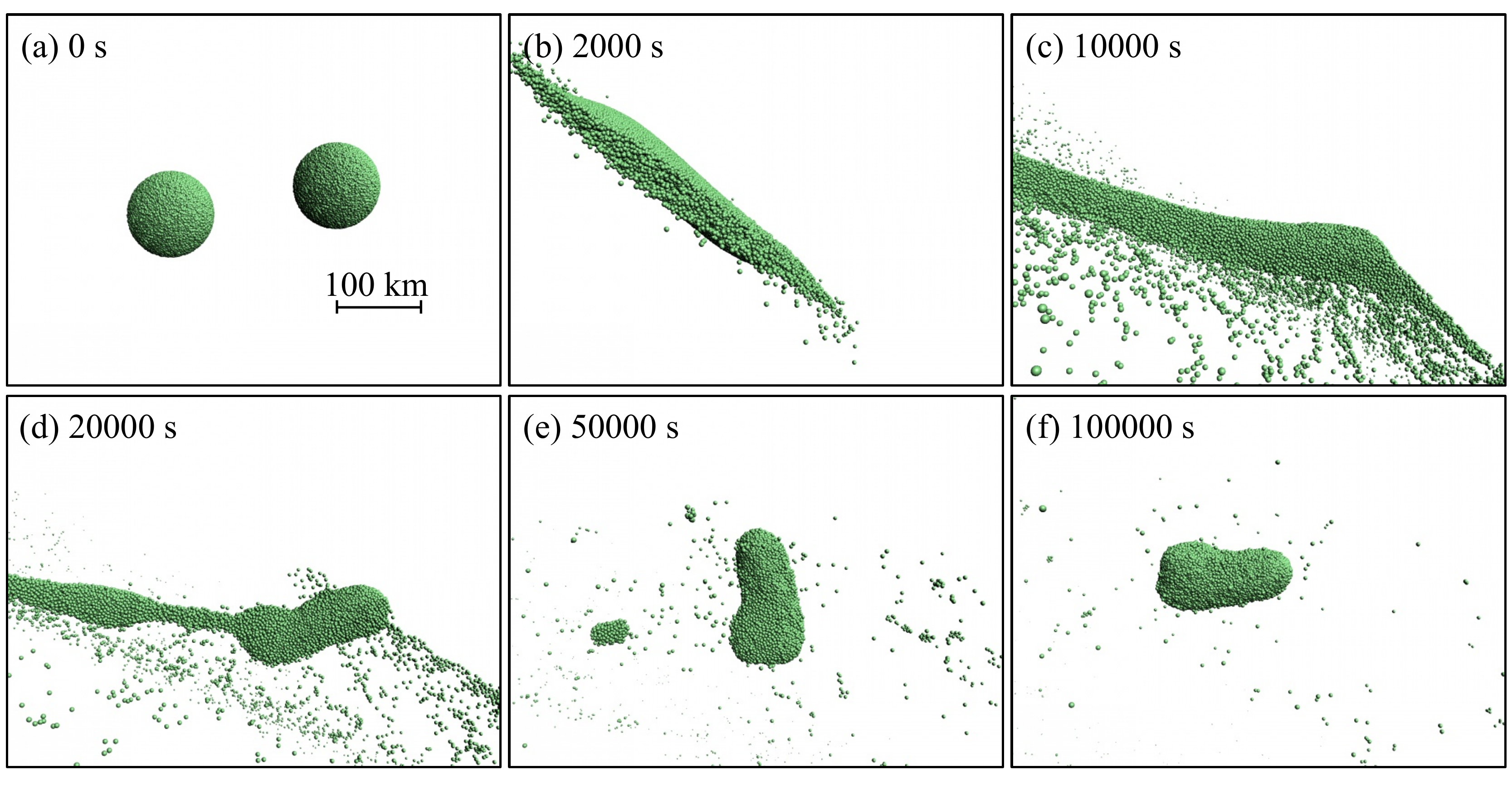}
    \caption{Snapshots of the impact simulation with $\theta_{{\rm imp}}=20\degr$ and $v_{{\rm imp}}=250\,{\rm m/s}$ at $0.0\,{\rm s}$(a), $2.0\times 10^{3}\,{\rm s}$(b), $1.0\times 10^{4}\,{\rm s}$(c), $2.0\times 10^{4}\,{\rm s}$(d), $5.0\times 10^{4}\,{\rm s}$(e), and $1.0\times 10^{5}\,{\rm s}$(f).}
    \label{subsequent-pictures-50-1.0-20-250}
  \end{center}
\end{figure}

\begin{figure}[!htb]
  \begin{center}
    \includegraphics[bb=0 0 960 767, width=1.0\linewidth,clip]{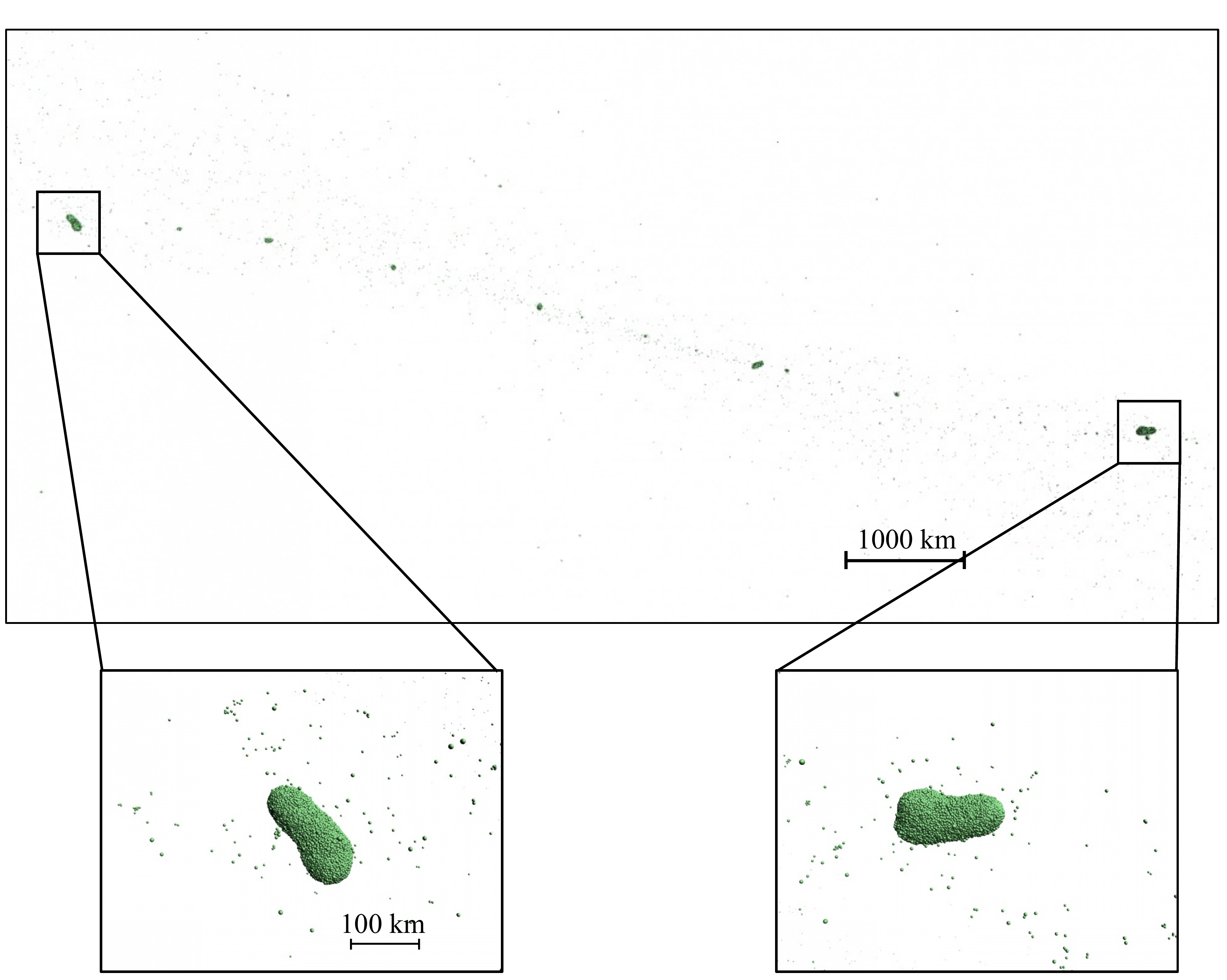}
    \caption{Zoom out view of the impact simulation with $v_{{\rm imp}}=250\,{\rm m/s}$ and $\theta_{{\rm imp}}=20\degr$ at $1.0\times 10^{5}\,{\rm s}$. Two enlarged figures represent the shape of the largest and second largest remnant, respectively.}
    \label{entire-picture-50-1.0-20-250-t=1.0e5}
  \end{center}
\end{figure}

Some hit-and-run collisions also produce elongated shapes. Fig.\,\ref{subsequent-pictures-50-1.0-20-250} shows snapshots of the impact with $v_{{\rm imp}}=250\,{\rm m/s}$ and $\theta_{{\rm imp}}=20\degr$, and Fig.\,\ref{entire-picture-50-1.0-20-250-t=1.0e5} shows a zoom out view of Fig.\,\ref{subsequent-pictures-50-1.0-20-250}f. The impact results in significant destruction and deformation (Fig.\,\ref{subsequent-pictures-50-1.0-20-250}b,c). Although two planetesimals do not merge (Fig.\,\ref{entire-picture-50-1.0-20-250-t=1.0e5}), the reaccretion of surrounding fragments produces two elongated shapes (Fig.\,\ref{subsequent-pictures-50-1.0-20-250}d-f and Fig.\,\ref{entire-picture-50-1.0-20-250-t=1.0e5}). Note that the largest and second largest objects in hit-and-run collisions have almost the same shape (Fig.\,\ref{entire-picture-50-1.0-20-250-t=1.0e5}).

\subsubsection{Hemispherical shapes}
\begin{figure}[!htb]
  \begin{center}
    \includegraphics[bb=0 0 960 512, width=1.0\linewidth,clip]{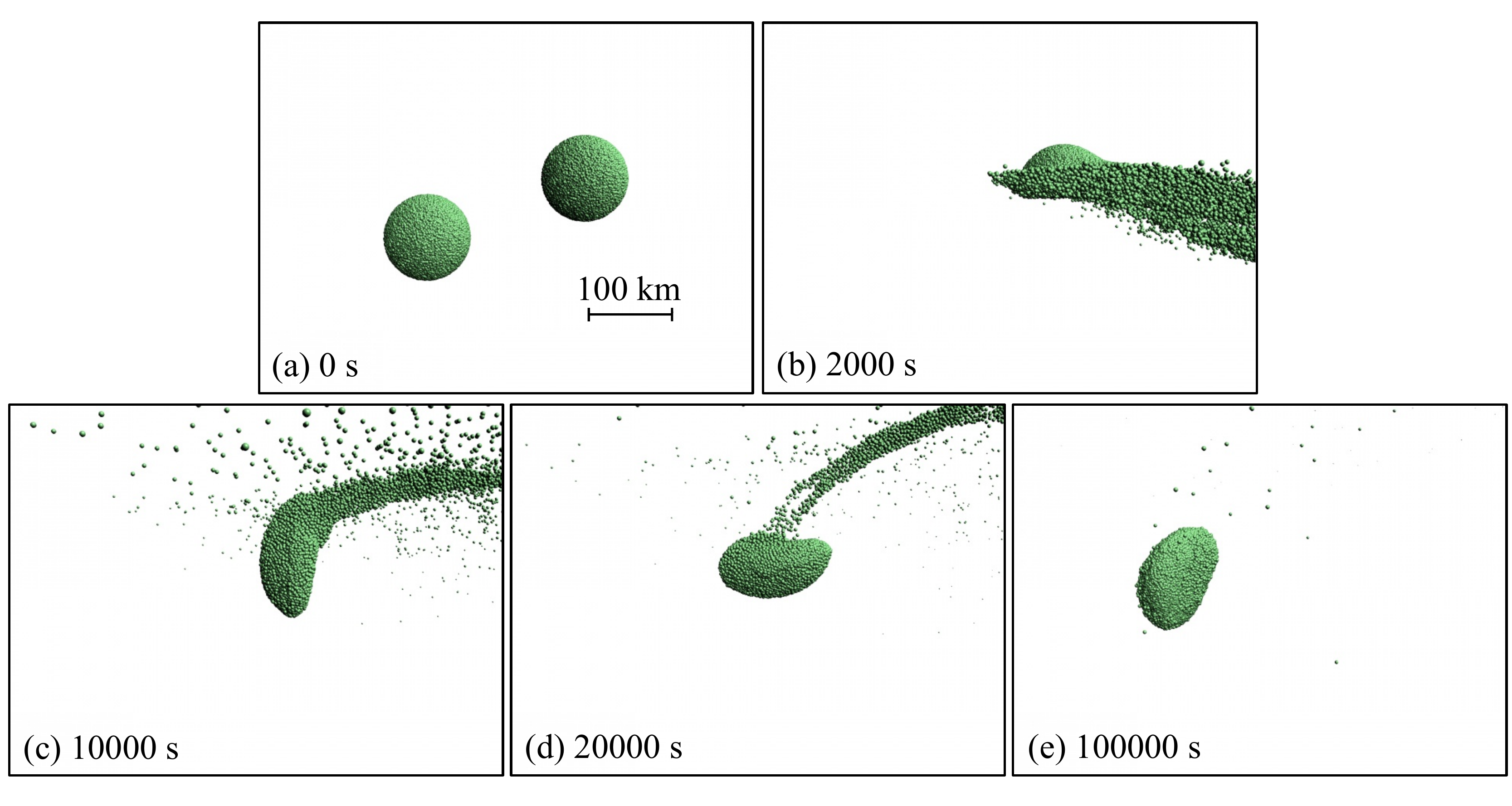}
    \caption{Snapshots of the impact simulation with $\theta_{{\rm imp}}=45\degr$ and $v_{{\rm imp}}=350\,{\rm m/s}$ at $0.0\,{\rm s}$(a), $2.0\times 10^{3}\,{\rm s}$(b), $1.0\times 10^{4}\,{\rm s}$(c), $2.0\times 10^{4}\,{\rm s}$(d), and $1.0\times 10^{5}\,{\rm s}$(e).}
    \label{subsequent-pictures-50-1.0-45-350}
  \end{center}
\end{figure}

In Fig.\,\ref{subsequent-pictures-50-1.0-45-350}, we show an impact forming hemispherical shapes. Significant destruction occurs around the impact point and a large amount of fragments is ejected straightforwardly (Fig.\,\ref{subsequent-pictures-50-1.0-45-350}b). This collisional truncation results in hemispherical shapes (Fig.\,\ref{subsequent-pictures-50-1.0-45-350}c-e).

\subsubsection{Super-catastrophic destruction}

\begin{figure}[!htb]
  \begin{center}
    \includegraphics[bb=0 0 960 498, width=1.0\linewidth,clip]{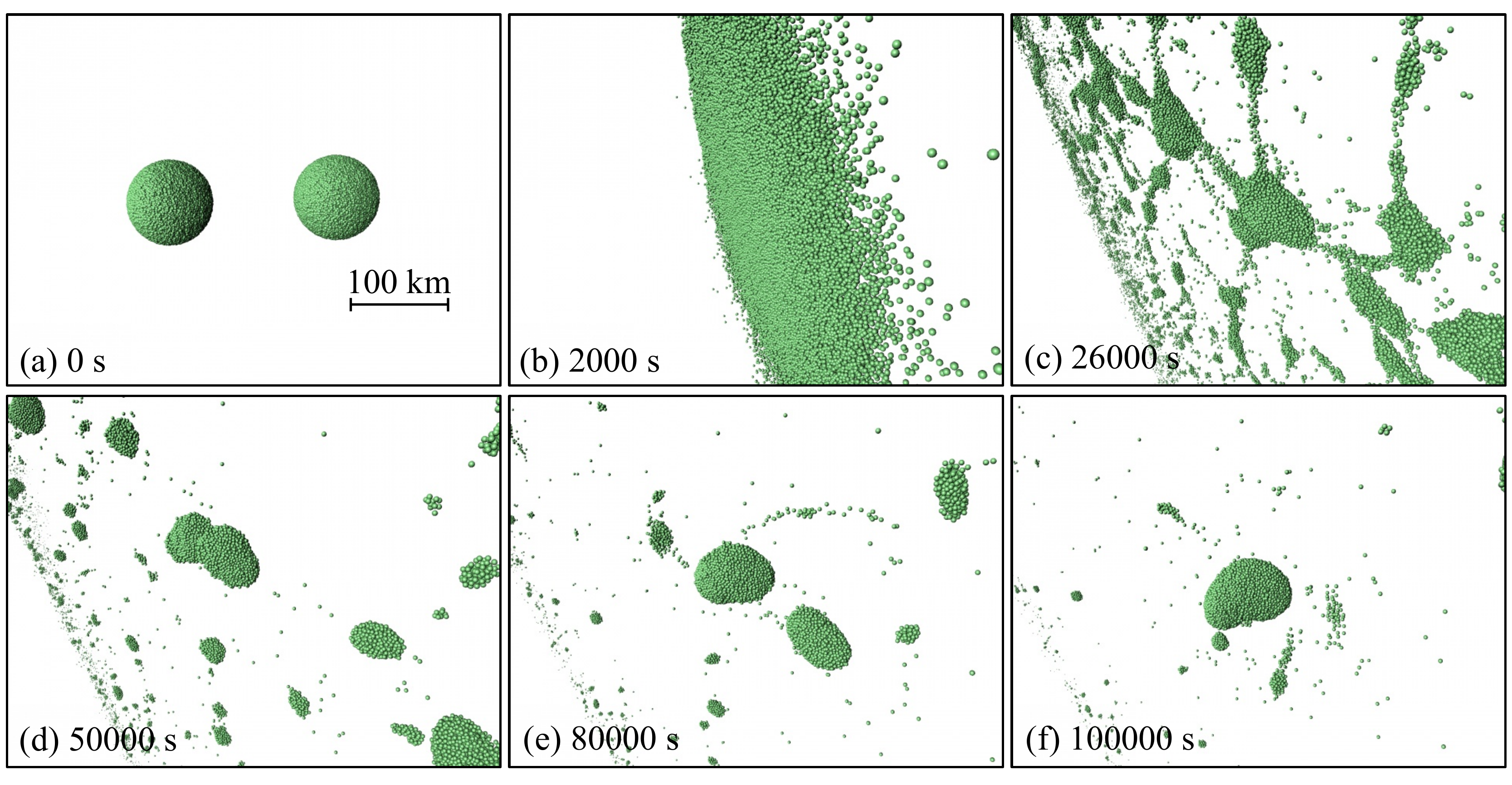}
    \caption{Snapshots of the impact simulation with $v_{{\rm imp}}=400\,{\rm m/s}$ and $\theta_{{\rm imp}}=5\degr$ at $0.0\,{\rm s}$(a), $2.0\times 10^{3}\,{\rm s}$(b), $2.6\times 10^{4}\,{\rm s}$(c), $5.0\times 10^{4}\,{\rm s}$(d), $8.0\times 10^{4}\,{\rm s}$(e), and $1.0\times 10^{5}\,{\rm s}$(f).}
    \label{subsequent-pictures-50-1.0-5-400}
  \end{center}
\end{figure}

Figure \ref{subsequent-pictures-50-1.0-5-400} represents the result of the impact simulation with $v_{{\rm imp}}=400\,{\rm m/s}$ and $\theta_{{\rm imp}}=5\degr$. The impact of very high $v_{{\rm imp}}$ produces a large curtain of ejected fragments (Fig.\,\ref{subsequent-pictures-50-1.0-5-400}b), and the gravitational fragmentation of the curtain forms many clumps (Fig.\,\ref{subsequent-pictures-50-1.0-5-400}c). Then the largest remnant is formed through the coalescence of clumps (Fig.\,\ref{subsequent-pictures-50-1.0-5-400}d-f).

In collisions with $M_{{\rm lr}} < 0.4M_{{\rm target}}$, the largest bodies are formed through significant reaccretion of ejecta. Even small difference of initial conditions produces significant difference of the distribution of ejecta, which leads to variety of shapes. Therefore, high-resolution simulations are required. We will conduct simulations with much higher number of SPH particles in our future work. In this paper, we just call impacts with $M_{{\rm lr}}<0.4M_{{\rm target}}$ super-catastrophic destruction, and do not classify shapes for such destructive impacts.

\subsection{Summary of shapes formed by collisions}

\begin{figure}[!htb]
 \begin{center}
   \includegraphics[bb=0 0 763 1080, width=0.8\linewidth,clip]{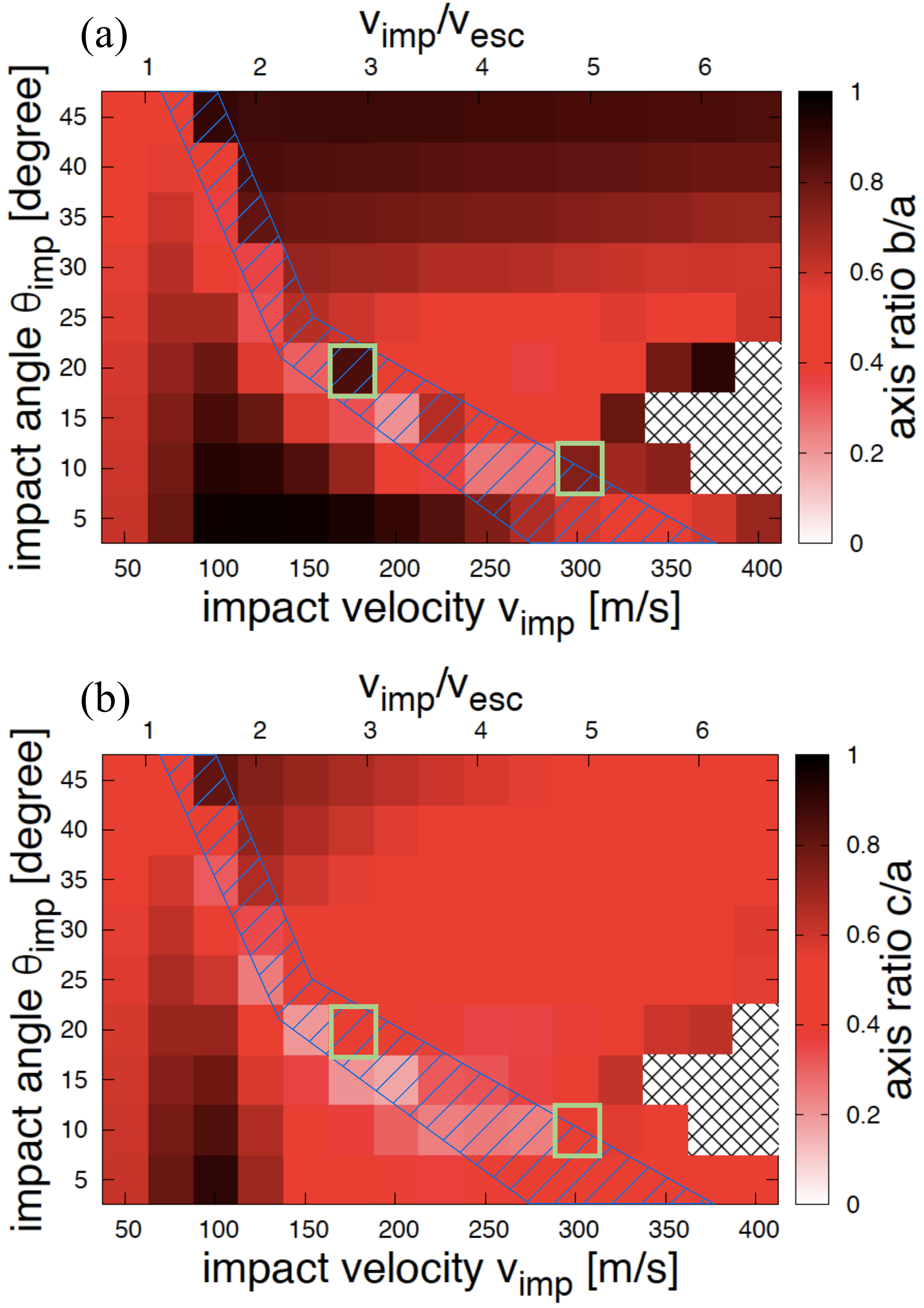}
   \caption{Dependence of the ratios $b/a$ and $c/a$ of the largest remnants on $v_{{\rm imp}}$ and $\theta_{{\rm imp}}$. Color represents (a) the ratio $b/a$, (b) the ratio $c/a$, respectively. For impacts in cross-hatched region, we do not measure the axis ratios of the largest remnants, because the mass of the largest remnants is too small (smaller than 0.15 $M_{{\rm target}}$). The meaning of the hatched regions is the same as in Fig.\,\ref{fmlr-contour-q=1.0-largest-only-detailed}. Parameters surrounded by green boxes represent impacts with the second collision as shown in Appendix C.}
   \label{axis-ratio-contour-q=1.0-detailed}
 \end{center}
\end{figure}

Figure \ref{axis-ratio-contour-q=1.0-detailed} shows the axis ratios of the largest remnants formed by impacts with various impact velocities and angles. For hit-and-run collisions, the largest and second largest bodies have similar shapes as shown in Fig.\,\ref{entire-picture-50-1.0-20-250-t=1.0e5}. Thus if the mass ratio of the first to second largest bodies is smaller than 2.0, we use the averaged values among two bodies for $b/a$ and $c/a$. Note that sharp variation of axis ratios at the hatched regions in Fig.\,\ref{axis-ratio-contour-q=1.0-detailed} is caused by the transition between merging and erosive or hit-and-run collisions. 

\begin{table}[!htb]
  \begin{center}
    \small
    \begin{tabular}{c c c c}\hline \hline
      Shape & \multicolumn{3}{c}{Thresholds}  \\
      \hline 
      Bilobed  & $b/a<0.7$ & $c/a<0.7$ & $M_{{\rm lr}}/M_{{\rm target}}=2.0$ \\ 
      Spherical & $b/a>0.7$ & $c/a>0.7$ & $\cdots$ \\ 
      Flat & $b/a>0.7$ & $c/a<0.7$ & $M_{{\rm lr}}/M_{{\rm target}}>1.0$ \\ 
      Elongated & $b/a<0.7$ & $c/a<0.7$ & $M_{{\rm lr}}/M_{{\rm target}}<2.0$ \\ 
      Hemispherical & $b/a>0.7$ & $c/a<0.7$ & $M_{{\rm lr}}/M_{{\rm target}}<1.0$ \\ 
      Super-catastrophic & $\cdots$ & $\cdots$ & $M_{{\rm lr}}/M_{{\rm target}}<0.4$ \\ \hline
    \end{tabular}
  \end{center}
  \caption{Thresholds of $b/a$, $c/a$, and $M_{{\rm lr}}/M_{{\rm target}}$ of the largest remnants for the categorization of shapes. All impacts with $M_{{\rm lr}}/M_{{\rm target}}<0.4$ are classified to super-catastrophic destruction regardless of the values of $b/a$ and $c/a$ of the largest remnants.}
  \label{table-categorization-threshold}
\end{table}

\begin{figure}[!htb]
 \begin{center}
 \includegraphics[bb=0 0 703 540, width=1.0\linewidth,clip]{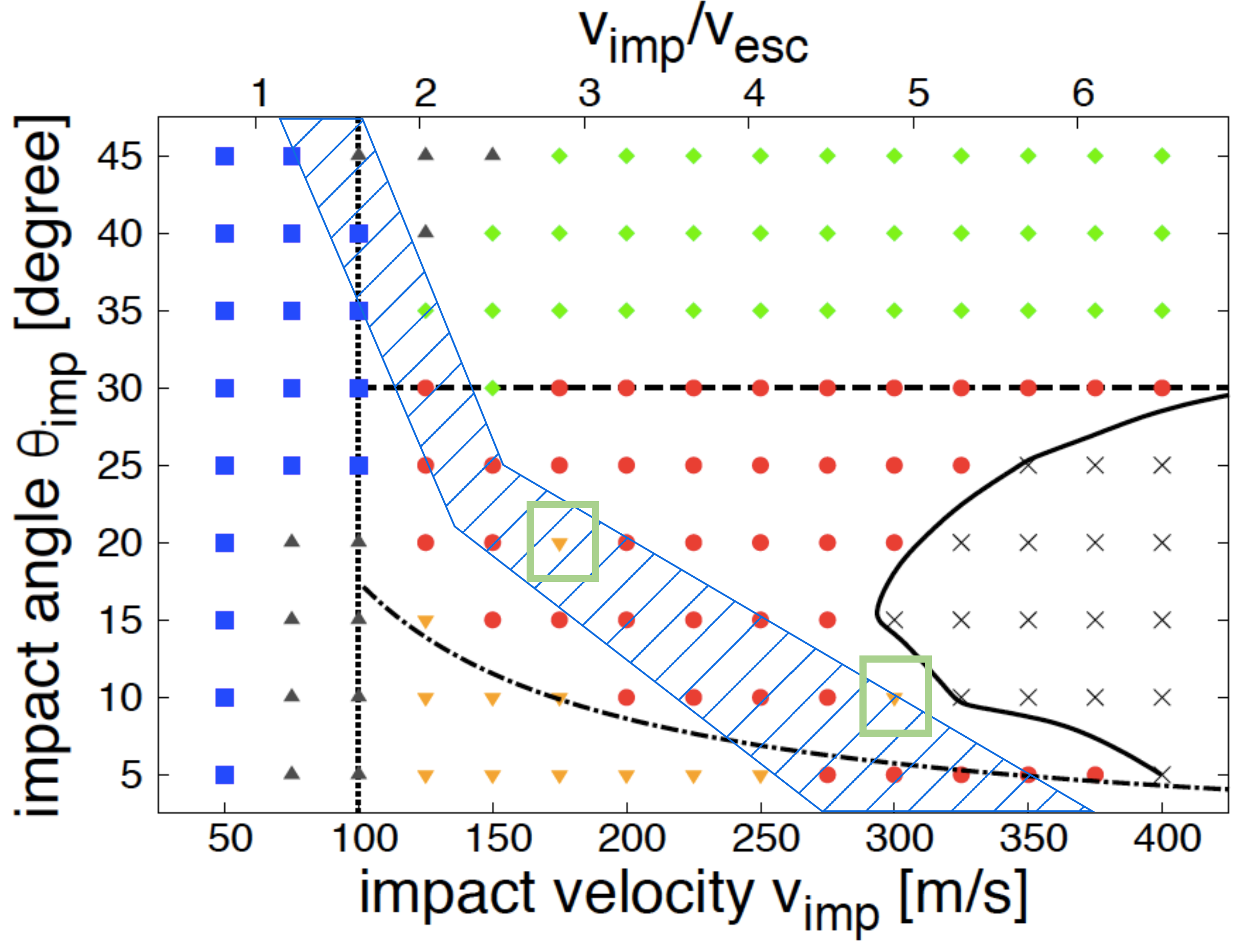}
 \caption{Summary of classification of resultant shapes. Blue squares represent the impact parameters, $v_{{\rm imp}}$ and $\theta_{{\rm imp}}$, producing bilobed shapes, gray triangles represent those of spherical shapes, orange inverted triangles represent those of flat shapes, red circles represent those of elongated shapes, light green diamonds represent those of hemispherical shapes, and black crosses represent those for super-catastrophic destruction. Dotted line shows $v_{{\rm imp}} = 1.6v_{{\rm esc}} = 100\,{\rm m/s}$, dashed line shows $\theta_{{\rm imp}} = 30\degr$, chain curve shows $v_{{\rm imp}}\sin \theta_{{\rm imp}} = 0.5v_{{\rm esc}} = 30\,{\rm m/s}$, and solid curve shows $M_{{\rm lr}} = 0.4M_{{\rm target}}$. The meanings of the hatched region and green boxes are the same as in Fig.\,\ref{fmlr-contour-q=1.0-largest-only-detailed} or Fig.\,\ref{axis-ratio-contour-q=1.0-detailed}.}
 \label{shape-category-division-and-symbol-q=1.0-detailed}
 \end{center}
\end{figure}

We categorize shapes of collisional outcomes into bilobed, spherical, flat, elongated, hemispherical, and super-catastrophic destruction as shown in Table \ref{table-categorization-threshold}. The classification given by Table \ref{table-categorization-threshold} mainly corresponds to the shapes formed via the processes shown in Section 4.3. Fig.\,\ref{shape-category-division-and-symbol-q=1.0-detailed} shows impact parameters producing the classified shapes, which indicates $v_{{\rm imp}}\sim 50\,{\rm m/s}$, or $v_{{\rm imp}}\sim 100\,{\rm m/s}$ and $\theta_{{\rm imp}} > 25\degr$ (bilobed shapes), $v_{{\rm imp}}\sim 100\,{\rm m/s}$ and $\theta_{{\rm imp}} < 25\degr$ (spherical shapes), $v_{{\rm imp}} > 100\,{\rm m/s}$ and $v_{{\rm imp}}\sin \theta_{{\rm imp}} < 30\,{\rm m/s}$ (flat shapes), $v_{{\rm imp}} > 100\,{\rm m/s}$, $v_{{\rm imp}}\sin \theta_{{\rm imp}} > 30\,{\rm m/s}$, and $\theta_{{\rm imp}} < 30\degr$ (elongated shapes), and $v_{{\rm imp}} > 100\,{\rm m/s}$ and $\theta_{{\rm imp}} > 30\degr$ (hemispherical shapes).

Note that two impacts resulting in flat shapes are in the elongated-shape region ($v_{{\rm imp}}=175\,{\rm m/s}, \theta_{{\rm imp}}=20\degr$ and $v_{{\rm imp}}=300\,{\rm m/s}, \theta_{{\rm imp}}=10\degr$). These impacts correspond to elongated-shape-forming collisions with the second collision as shown in Appendix C. We consider these impacts as elongated-shape-forming collisions from the shapes observed in the simulations.

Based on the classification in Fig\,\ref{shape-category-division-and-symbol-q=1.0-detailed}, we find the specific conditions to determine the shapes, which are written as

\begin{itemize}
\item bilobed and spherical shapes: $v_{{\rm imp}}<1.6v_{{\rm esc}}$,
\item flat shapes: $v_{{\rm imp}}>1.6v_{{\rm esc}}$ and $v_{{\rm imp}}\sin \theta_{{\rm imp}}<0.5v_{{\rm esc}}$,
\item hemispherical shapes: $v_{{\rm imp}}>1.6v_{{\rm esc}}$ and $\theta_{{\rm imp}}>30\degr$,
\end{itemize}

\noindent and

\begin{itemize}
\item elongated shapes: $v_{{\rm imp}}>1.6v_{{\rm esc}}$, $\theta_{{\rm imp}}<30\degr$, $v_{{\rm imp}}\sin \theta_{{\rm imp}}>0.5v_{{\rm esc}}$, and $M_{{\rm lr}}>0.4M_{{\rm target}}$,
\end{itemize}

\noindent where two-body escape velocity $v_{{\rm esc}}\approx 60\,{\rm m/s}$. The hatched region in Fig.\,\ref{shape-category-division-and-symbol-q=1.0-detailed} divides the elongated-shape region into two parts. Elongated shapes formed with parameters in the left part of the hatched region are formed by merging collisions (see Fig.\,\ref{subsequent-pictures-50-1.0-15-200-100000}), and those in the right part are formed by hit-and-run collisions (see Fig.\,\ref{subsequent-pictures-50-1.0-20-250}). 

\section{Discussion}
\subsection{Four conditions required for the formation of elongated shapes}
The threshold of $v_{{\rm imp}} > 1.6v_{{\rm esc}}$ is required for significant deformation. We estimate necessary impact velocity to deform planetesimals as follows. Frictional force of $\mu_{d}p$ acts on the area of $\sim \pi R_{t}^{2}$ and the energy dissipation occurs due to frictional deformation on the length scale of $\sim 4R_{t}$. The dissipated energy $E_{{\rm dis}}$ is estimated as

\begin{equation}
  E_{{\rm dis}}=4\pi R_{t}^{3}\mu_{d}p.
  \label{critical-dissipated-energy}
\end{equation}

\noindent The timescale for deformation of whole bodies is estimated to be $2R_{t}/v_{{\rm imp}}$, which is much longer than the shock passing time $\sim 2R_{t}/C_{s}$, where $C_{s}\approx 3\,{\rm km/s}$ is the sound speed. High pressure states caused by shocks are already relaxed before the end of the deformation, and shocks do not contribute to the pressure for frictional force given in Eq.\,(\ref{critical-dissipated-energy}). Therefore, the pressure is mainly determined by the self-gravity and estimated to be central pressure of a planetesimal with the radius of $R_{t}$ and uniform density of $\rho_{0}$, given by

\begin{equation}
  p=\frac{2}{3}\pi G \rho_{0}^{2} R_{t}^{2}.
  \label{typical-central-pressure}
\end{equation}

\noindent Equating the total initial kinetic energy for two equal-mass bodies $(1/4)M_{{\rm target}}v_{{\rm imp}}^{2}$ to $E_{{\rm dis}}$, we obtain the critical deformation velocity as

\begin{align}
  v_{{\rm imp,crit}} & =\sqrt{4E_{{\rm dis}}/M_{{\rm target}}} = \sqrt{3\mu_{d}}v_{{\rm esc}} \nonumber \\ &= 1.587\Bigl( \frac{\mu_{d}}{\tan (40\degr)} \Bigr) v_{{\rm esc}}.
  \label{critical-impact-velocity}
\end{align}

\noindent The impact velocity obtained in Eq.\,(\ref{critical-impact-velocity}) well agrees with $v_{{\rm imp}} = 1.6v_{{\rm esc}}$ in spite of rough estimation of the dissipated energy $E_{{\rm dis}}$. 

The condition of $\theta_{{\rm imp}}<30\degr$ is needed for the avoidance of hemispherical shapes caused by hit-and-run collisions. For $\theta_{{\rm imp}} \geq 30\degr$ half or smaller of a target is directly interacted by an impactor, resulting in hit-and-run collisions (\citealt{Asphaug2010,Leinhardt-and-Stewart2012}). To form elongated shapes, it is necessary for almost whole volume of two planetesimals to be deformed. For $\theta_{{\rm imp}}<30\degr$ most of two planetesimals are directly affected by collisions, which leads to deformation to be elongated shapes. 

Collisional elongation requires large shear velocity $v_{{\rm imp}}\sin \theta_{{\rm imp}}>0.5v_{{\rm esc}}$, while impacts with $v_{{\rm imp}}\sin \theta_{{\rm imp}}<0.5v_{{\rm esc}}$ produce flat shapes. Elongated shapes are formed through stretch of planetesimals in the direction of shear velocity (see Fig.\,\ref{subsequent-pictures-50-1.0-15-200-100000},\,\ref{subsequent-pictures-50-1.0-20-250}). However, the self-gravity prevents deformation, which occurs if $v_{{\rm imp}}\sin \theta_{{\rm imp}}\ll v_{{\rm esc}}$. We find that elongation needs $v_{{\rm imp}}\sin \theta_{{\rm imp}} > v_{{\rm esc}}/2$.

Super-catastrophic destruction with $M_{{\rm lr}}<0.4M_{{\rm target}}$ produces many small remnants, which mainly have spherical shapes as shown in Fig.\,\ref{subsequent-pictures-50-1.0-5-400}. Elongated shapes may not be formed through super-catastrophic destruction. Thus for certain formation of elongated shapes, $M_{{\rm lr}}>0.4M_{{\rm target}}$ is required.

Distribution of pressure is determined by the self-gravity unless the impact velocity is comparable to or larger than the sound speed. Since frictional force is proportional to the pressure, the friction is also determined by the self-gravity. Thus, unless the material strength is dominant, force on bodies (right hand side of Eq.\,(\ref{EoM})) is solely determined by the self-gravity, so that results of impacts are characterized by dimensionless velocity $v_{{\rm imp}}/v_{{\rm esc}}$ regardless of the scale or size of planetesimals. For rocky planetesimals with $R_{t} \geq 0.5\,{\rm km}$, the friction is dominant rather than the material strength. For $R_{t} \leq 200\,{\rm km}$, $v_{{\rm esc}}$ is smaller than $0.1 C_{s}$. Therefore, the conditions to form elongated shapes and the shape classification of Fig.\,\ref{shape-category-division-and-symbol-q=1.0-detailed} with upper horizontal axis are also valid for equal-mass impacts with the angle of friction of $40\degr$ and $10^{0}\,{\rm km} \la R_{t} \la 10^{2}\,{\rm km}$. 

\subsection{Applications}
We analyze shapes of 139 asteroids with diameters $D > 80\,{\rm km}$, which are obtained using DAMIT database (http://astro.troja.mff.cuni.cz/projects/asteroids3D/web.php). We derive axis ratios of asteroids in DAMIT according to experimental method (\citealt{Fujiwara-et-al1978}). We find 20 irregular shaped asteroids that have $c/a < 0.6$ and $D > 80\,{\rm km}$. These irregular shaped asteroids include elongated ones with $b/a < 0.6$ ((63) Ausonia, (216) Kleopatra, (624) Hektor) and flat ones with $b/a > 0.9$ and $c/a < 0.5$ ((419) Aurelia, (471) Papagena). Therefore, $\sim 10\%$ of asteroids with $D > 80\,{\rm km}$ have irregular shapes. Note that this fraction may become smaller because DAMIT preferentially contains irregular shaped asteroids due to the light-curve inversion technique. However, in spite of the error, DAMIT seems to accurately measure $b/a$ for asteroids with $b/a<0.6$. For example, $b/a$ of asteroid Itokawa obtained from the light curve is 0.5 (\citealt{Kaasalainen-et-al2003}), which is similar value to $b/a = 0.55$ obtained from the in-situ observation (\citealt{Fujiwara-et-al2006}). Therefore, we discuss the statistics of asteroidal shapes based on DAMIT.

(624) Hektor is a Jupiter trojan, and the others are main-belt asteroids. In the main belt, the keplerian velocity is $v_{{\rm K}} \approx 20\,{\rm km/s}$ and the mean impact velocity is roughly estimated to be $\sqrt{e^{2}_{{\rm ave}}+i^{2}_{{\rm ave}}}v_{{\rm K}} \approx 4\,{\rm km/s}$ with the mean orbital eccentricity of $e_{{\rm ave}} = 0.15$ and inclination of $i_{{\rm ave}} = 0.13$ (\citealt{Ueda-et-al2017}). This mean impact velocity is much higher than that treated in our simulations ($v_{{\rm imp}}<400\,{\rm m/s}$). We estimate distribution of impact velocities between main-belt asteroids using orbital parameters obtained from the JPL small-body Database Search Engine (https://ssd.jpl.nasa.gov/sbdb\_query.cgi\#x) and the method to obtain the relative velocity at the orbital crossing according to \citet{Kobayashi-and-Ida2001}. This gives the mean collisional velocity of $5\,{\rm km/s}$. The probability that impact velocities become less than $400\,{\rm m/s}$ is about 0.15\%. Therefore, expected production of irregular shaped asteroids due to low-velocity impacts in the present solar system is too low to reproduce the current fraction of the irregular shaped asteroids. 

Similar-mass impacts with the mean impact velocity in the main belt result in catastrophic destruction, which also produces irregular shapes and asteroid families. An irregular shaped asteroid formed through a recent catastrophic destruction may be contained in an asteroid family. According to AstDyS-2 database (http://hamilton.dm.unipi.it/astdys/), among 20 irregular shaped asteroids with $D > 80\,{\rm km}$ that we find in DAMIT, only three asteroids ((20) Massalia, (63) Ausonia, (624) Hektor) are contained in asteroid families. However, (20) Massalia and (624) Hektor are the largest remnants of asteroid families that are formed through cratering impacts (\citealt{Vokrouhlicky-et-al2006,Rozehnal-et-al2016}). Thus 19 out of these 20 irregular shaped asteroids are probably not formed through recent catastrophic destruction: Catastrophic destruction is the minor process for the formation of irregular shapes of large asteroids, which is consistent with their collisional lifetimes estimated to be much longer than the age of the solar system (\citealt{Obrien-and-Greenberg2005}).

Bilobed shapes are also formed through largely destructive impacts. Many remnants are formed in a large curtain of ejected fragments as shown in Fig.\,\ref{subsequent-pictures-50-1.0-5-400}, and then these remnants may again collide with each other with $v_{{\rm imp}}\sim v_{{\rm esc}}$, which leads to the formation of bilobed shapes. We additionally conduct a higher resolution simulation of a largely destructive impact, which shows that bilobed asteroids are formed. However, flat shapes are not formed in the destructive impact. 

For large-mass-ratio impacts, deformation occurs only in the scale of impactors much smaller than that of targets, so that overall deformation does not occur via a single collision. Our additional simulations with impactor-to-target mass ratio 1/64 show that non-disruptive impacts ($M_{{\rm lr}}>0.5M_{{\rm target}}$) with various impact velocities of $v_{{\rm imp}} = 500 - 3000\,{\rm m/s}$ and angles of $\theta_{{\rm imp}} = 0 - 40\degr$ do not form irregular shapes with $c/a \la 0.7$. Although such impacts are frequent, isotropic impacts to almost spherical asteroids do not form irregular shapes. 

Therefore, irregular shapes, especially flat shapes, of asteroies with $D > 80\,{\rm km}$ are likely to be formed through similar-mass and low-velocity impacts, which are unlikely to occur in the present solar system. Relative velocities between planetesimals are increased by gravitational interaction with planets, especially Jupiter (e.g., \citealt{Kobayashi-et-al2010}). Thus, collisional velocities in the main belt may be much lower prior to Jupiter formation. Jupiter formation may significantly deplete asteroids (\citealt{Bottke-et-al2005}), and similar-mass impacts may be frequent prior to Jupiter formation. Irregular shaped asteroids are possibly formed in such environments. Therefore, irregular shaped asteroids with $D > 80\,{\rm km}$ may be formed in the primordial environment and remain the same until today. 

\section{Summary}
 Asteroids are believed to be remnants of planetesimals formed in planet formation era and may retain information of the history of the solar system. Irregular shapes of asteroids are possible to be formed through collisional destruction and coalescence of planetesimals. Thus clarifying impact conditions to form specific shapes of asteroids leads to constrain epoch or collisional environment forming those asteroids.

Our simulations show the relationship between impact conditions and resultant shapes of planetesimals. We carry out simulations of collisions between planetesimals using SPH method for elastic dynamics with the self-gravity and the models for fracture and friction. We consider collisions between two planetesimals with the radius of $50\,{\rm km}$, because significant shape deformation occurs in equal-mass impacts. We vary the impact velocity $v_{{\rm imp}}$ from $50\,{\rm m/s}$ to $400\,{\rm m/s}$ and the impact angle $\theta_{{\rm imp}}$ from $5\degr$ to $45\degr$. Then we measure shape of the largest remnant formed in each impact simulation.

We confirm that various shapes are formed by equal-mass impacts. We classify shapes of the largest remnants into 5 categories if the mass of the largest remnant $M_{{\rm lr}}$ is larger than 0.4 of that of an initial planetesimal $M_{{\rm target}}$.  The result of the shape classification is as follows:

\begin{itemize}
\item For $v_{{\rm imp}} \sim 50\,{\rm m/s}$, or $v_{{\rm imp}}\sim 100\,{\rm m/s}$ and $\theta_{{\rm imp}} > 25\degr$, bilobed shapes are formed because of merging of planetesimals with preserving the initial spherical shapes (see Fig.\,\ref{subsequent-pictures-50-1.0-30-50}).
\item For $v_{{\rm imp}} \sim 100\,{\rm m/s}$ and $\theta_{{\rm imp}} < 20\degr$, spherical shapes are formed because a part of planetesimal is crushed (see Fig.\,\ref{subsequent-pictures-50-1.0-10-100}).
\item For $v_{{\rm imp}}>100\,{\rm m/s}$ and $v_{{\rm imp}}\sin \theta_{{\rm imp}}<30\,{\rm m/s}$, flat shapes are formed because of nearly head-on collisions and larger deformation (see Fig.\,\ref{subsequent-pictures-50-1.0-5-200}).
\item For $v_{{\rm imp}}>100\,{\rm m/s}$, $v_{{\rm imp}}\sin \theta_{{\rm imp}}>30\,{\rm m/s}$, and $\theta_{{\rm imp}} < 30\degr$, elongated shapes are formed because planetesimals are stretched to the direction perpendicular to the line joining centers of two contacting planetesimals (see Fig.\,\ref{subsequent-pictures-50-1.0-15-200-100000}).
\item For $v_{{\rm imp}} > 100\,{\rm m/s}$ and $\theta_{{\rm imp}}>30\degr$, hemispherical shapes are formed because of excavation of edges of planetesimals (see Fig.\,\ref{subsequent-pictures-50-1.0-45-350}).
\end{itemize}

As a result of the shape classification, we find four conditions to form elongated shapes with the ratio $b/a$ smaller than 0.7. Those four conditions and the meaning of each condition are as follows:

\begin{itemize}
\item $v_{{\rm imp}} > 1.6v_{{\rm esc}}$: Overall deformation of planetesimals requires large impact velocity.
\item $\theta_{{\rm imp}} < 30\degr$: Impacts with large impact angles result in erosion of only edges of planetesimals.
\item $v_{{\rm imp}}\sin \theta_{{\rm imp}} > 0.5v_{{\rm esc}}$: Elongated shapes are formed through stretch of planetesimals to the direction of shear velocity $v_{{\rm imp}}\sin \theta_{{\rm imp}}$, so that large shear velocity is also required.
\item $M_{{\rm lr}}>0.4M_{{\rm target}}$: In largely destructive impacts the largest remnants are formed through violent reaccumulation of fragments and resultant shapes tend to be spherical.
\end{itemize}

\noindent As we discussed in Section 5.1, these conditions are also valid for equal-mass impacts with the angle of friction of $40\degr$ and $10^{0}\,{\rm km} \la R_{t} \la 10^{2}\,{\rm km}$ although we have not yet confirmed this through numerical experiments.

According to our simulations, various irregular shapes are formed through impacts with two equal-mass planetesimals and low impact velocities $< 400\,{\rm m/s}$. Impacts with the average relative velocity in the main belt $\approx 5\,{\rm km/s}$ mainly result in catastrophic destruction for similar-mass impacts or moderate destruction for high-mass-ratio impacts. However, both catastrophic destruction and impacts with high mass ratio are unlikely to produce flat shapes as we discussed in Section 5.2 based on our additional simulations. Asteroids with diameters $> 80\,{\rm km}$ have longer collisional lifetimes than the age of the solar system. Therefore, we suggest that irregular shapes, especially flat shapes, of asteroids with diameters $> 80\,{\rm km}$ are likely to be formed through similar-mass and low-velocity impacts in the primordial environment prior to the formation of Jupiter.

We only consider collisions between two rocky planetesimals with the radius of $50\,{\rm km}$ and the limited impact velocity range. Thus investigations of following impacts are our future work: impacts with different radii of planetesimals, high mass ratios of colliding two planetesimals, and higher impact velocities, which form smaller fragments. Clarifying more detailed relationship between impact conditions and resultant shapes of planetesimals may allow us to extract more detailed information of the history of the solar system from shapes of asteroids.

\begin{acknowledgement}
  We thank Hidekazu Tanaka and Ryuki Hyodo for the useful discussions and comments, and Martin Jutzi and Natsuki Hosono for giving us the useful information about the numerical simulation methods. We also thank an anonymous reviewer for valuable and detailed comments on our manuscript. KS is supported by JSPS KAKENHI Grant Number JP 17J01703. HK and SI are supported by Grant-in-Aid for Scientific Research (18H05436, 18H05437, 18H05438, 17K05632, 17H01105, 17H01103, 23244027, 16H02160). Numerical simulations in this work were carried out on Cray XC30 at Center for Computational Astrophysics, National Astronomical Observatory of Japan.
\end{acknowledgement}

\bibliography{mybibfile}

\begin{appendix}
\section{Time development method}
Here, we describe detailed equation of the leapfrog integrator used in this study. At the $n$-th step we update the position of the $i$-th particle $\bm{x}_{i}^{n}$ and other quantities of the $i$-th particle $q_{i}^{n}$ as

\begin{align}
&\bm{x}_{i}^{n+1}=\bm{x}_{i}^{n}+\bm{v}_{i}^{n}\Delta t + \frac{1}{2}\Bigl( \frac{d\bm{v}_{i}}{dt}\Bigr)^{n}\Delta t^{2},\nonumber \\
&q_{i}^{n+1}=q_{i}^{n}+\frac{1}{2}\Bigl[ \Bigl( \frac{dq_{i}}{dt} \Bigr)^{n} + \Bigl( \frac{dq_{i}}{dt} \Bigr)^{n+1} \Bigr]\Delta t, \tag{A1} \label{leapfrog-method}
\end{align}

\noindent where $\Delta t$ is the timestep.

In Eq.\,(\ref{leapfrog-method}), for example $\bm{v}_{i}^{n+1}$ is determined by $(d\bm{v}_{i}/dt)^{n+1}$. However, to calculate $(d\bm{v}_{i}/dt)^{n+1}$ with the equation of motion (\ref{EoM}) we need $\bm{v}_{i}^{n+1}$, so that we cannot directly derive $\bm{v}_{i}^{n+1}$. Thus we update physical quantities as following procedure: Firstly, we predict the quantities of the ($n+1$)-th step only using the quantities of the $n$-th step as

\begin{equation}
q_{i}^{\ast}=q_{i}^{n}+\Bigl( \frac{dq_{i}}{dt} \Bigl)^{n}\Delta t. \tag{A2}
\label{leapfrog-trick}
\end{equation}

\noindent At the same time we update the positions as Eq.\,(\ref{leapfrog-method}). Then we calculate $(dq_{i}/dt)^{n+1}$ using $q_{i}^{\ast}$ and $\bm{x}_{i}^{n+1}$, and we obtain $q_{i}^{n+1}$ from Eq.\,(\ref{leapfrog-method}). We can reuse $(dq_{i}/dt)^{n+1}$ at the next step, so that we calculate derivatives of variables only once at every step. Moreover, if all variables vary linearly in time, this procedure does not produce any integration error. Therefore, this integration scheme has second-order accuracy in time.

The timestep $\Delta t$ is determined by considering the Courant condition as

\begin{equation}
\Delta t = \min_{i} C_{{\rm CFL}}\frac{h_{i}}{C_{s,i}}, \tag{A3}
\label{timestep}
\end{equation}

\noindent where $C_{s,i}$ is local bulk sound speed at the position of the $i$-th particle. $C_{s,i}$ is calculated from the equation of state, the density, the specific internal energy, and the pressure of the $i$-th particle. We adopt the value of $C_{{\rm CFL}}=0.5$.

\section{Measurement of axis lengths of the largest remnants}
The inertia moment tensor of the largest remnant composed of $k$ SPH particles is calculated as

\begin{equation}
  I^{\alpha \beta}=\sum_{k}m_{k}\Bigl[ (x_{k}^{\gamma}-x_{{\rm CoM}}^{\gamma})(x_{k}^{\gamma}-x_{{\rm CoM}}^{\gamma})\delta^{\alpha \beta} - (x_{k}^{\alpha}-x_{{\rm CoM}}^{\alpha})(x_{k}^{\beta}-x_{{\rm CoM}}^{\beta}) \Bigr],
  \label{inertia-moment-tensor}
\end{equation}

\noindent where $\bm{x}_{{\rm CoM}}$ is the position vector at the center of mass of the largest remnant. Then, three principal moments of inertia $I_{1}$, $I_{2}$, and $I_{3}$ are obtained from $I^{\alpha \beta}$. Here, $I_{1}>I_{2}>I_{3}$. For a uniform ellipsoid with the length of major axis $a$, intermediate axis $b$, and minor axis $c$, the three principal moments of inertia are represented as

\begin{align}
  &I_{1}=\frac{1}{20}(a^{2}+b^{2})M_{{\rm lr}}, \nonumber \\
  &I_{2}=\frac{1}{20}(a^{2}+c^{2})M_{{\rm lr}}, \nonumber \\
  &I_{3}=\frac{1}{20}(b^{2}+c^{2})M_{{\rm lr}}. \label{principal-moment-of-inertia-of-ellipsoid}
\end{align}

\noindent Eq.\,(\ref{principal-moment-of-inertia-of-ellipsoid}) is rewritten as

\begin{align}
  &a=\sqrt{\frac{10(I_{1}+I_{2}-I_{3})}{M_{{\rm lr}}}}, \nonumber \\
  &b=\sqrt{\frac{10(I_{1}-I_{2}+I_{3})}{M_{{\rm lr}}}}, \nonumber \\
  &c=\sqrt{\frac{10(-I_{1}+I_{2}+I_{3})}{M_{{\rm lr}}}}. \label{axis-lengths-using-inertia-moment}
\end{align}

\noindent Therefore, we obtain $I_{1}$, $I_{2}$, $I_{3}$, and $M_{{\rm lr}}$ of the largest remnant through a simulation, and then derive its $a$, $b$, and $c$ from Eq.\,(\ref{axis-lengths-using-inertia-moment}).

\section{Elongated-shape-forming impacts with the second collisions}
\begin{figure}[!htb]
  \begin{center}
    \includegraphics[bb=0 0 960 498, width=1.0\linewidth,clip]{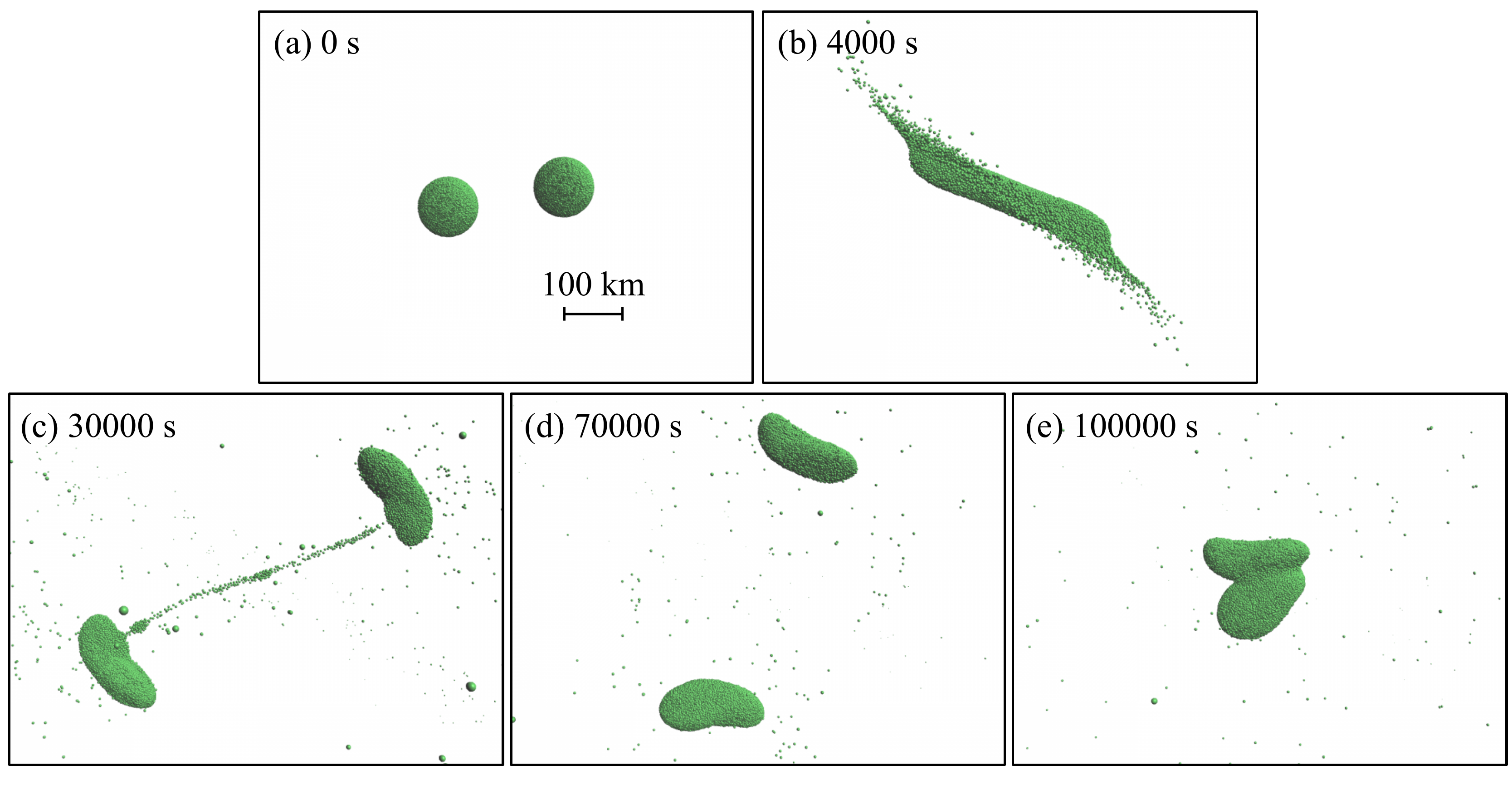}
    \caption{Snapshots of the impact simulation with $v_{{\rm imp}}=175\,{\rm m/s}$ and $\theta_{{\rm imp}}=20\degr$ at $0.0\,{\rm s}$(a), $4.0\times 10^{3}\,{\rm s}$(b), $3.0\times 10^{4}\,{\rm s}$(c), $7.0\times 10^{4}\,{\rm s}$(d), and $1.0\times 10^{5}\,{\rm s}$(e).}
    \label{subsequent-pictures-50-1.0-20-175}
  \end{center}
\end{figure}

Figure \ref{subsequent-pictures-50-1.0-20-175} represents a collision with $v_{{\rm imp}}=175\,{\rm m/s}$ and $\theta_{{\rm imp}}=20\degr$, which results in the second collision of two elongated objects. As in Fig\,\ref{subsequent-pictures-50-1.0-15-200-100000}, the first collision produces two elongated objects (Fig\,\ref{subsequent-pictures-50-1.0-20-175}b,c). However, the energy dissipation by the collision makes colliding bodies gravitationally bounded (Fig.\,\ref{subsequent-pictures-50-1.0-20-175}d), and the resultant body formed by the merging is no longer elongated object (Fig.\,\ref{subsequent-pictures-50-1.0-20-175}e). Although the resultant body is not elongated, this impact should be also classified to elongated-shape-forming collision. 

\end{appendix}

\end{document}